\documentclass[aps, pre, onecolumn, amsmath, superscriptaddress,showkeys,showpacs]{revtex4-1}
\usepackage{graphicx}
\usepackage{dcolumn}
\usepackage{bm}
\usepackage[utf8]{inputenc}
\usepackage[T1]{fontenc}
\usepackage{mathptmx}
\usepackage{amsmath, amssymb, amsthm}
\usepackage{float}
\usepackage{pgf}
\usepackage{color}
\usepackage{epstopdf}
\usepackage{hyperref}
\usepackage{lipsum}
\usepackage{soul}
\usepackage{ulem}
\usepackage{float}
\usepackage{wrapfig}
\hypersetup{
	colorlinks = true,
	citecolor=blue,
	runcolor=red,
	linkcolor=red,
	linkbordercolor = {white},}

\newcommand{\includegraphic}[5][,]{
	\setbox1=\hbox{\includegraphics[#1]{#2}}
	\leavevmode\rlap{\usebox1}
	\rlap{\hspace*{#4}\raisebox{\dimexpr\ht1-#5\baselineskip}{\normalsize{#3}}}
	\phantom{\usebox1}}
\begin{document}
	
	
	\title{Infection spreading and recovery in a square lattice}
	
	\author{Suman Saha}
	\affiliation{Centre for Mathematical Biology and Ecology, Department of Mathematics, Jadavpur University, Kolkata 700032, India.}
	\affiliation{Translational Health Science and Technology Institute, NCR Biotech Science Cluster 3rd Milestone,  Faridabad 121001, India.}
	\author{Arindam Mishra}
	\affiliation{Centre for Mathematical Biology and Ecology, Department of Mathematics, Jadavpur University, Kolkata 700032, India.}
	\author{Syamal K. Dana}
	\affiliation{Centre for Mathematical Biology and Ecology, Department of Mathematics, Jadavpur University, Kolkata 700032, India.}
	\author{Chittaranjan Hens}
	\affiliation{Physics and Applied Mathematics Unit, Indian Statistical Institute, Kolkata 700108, India}
	\author{Nandadulal Bairagi}
	\email{nbairagi.math@jadavpuruniversity.in}
	\affiliation{Centre for Mathematical Biology and Ecology, Department of Mathematics, Jadavpur University, Kolkata 700032, India.}

	\begin{abstract}
		We investigate spreading and recovery of disease in a square lattice, and in particular, emphasize the role of the initial distribution of infected patches in the network, on the progression of an endemic and initiation of a  recovery process, if any, due to migration of both the susceptible and infected hosts. 
		The disease starts in the lattice with three possible initial distribution patterns of infected  and infection-free
		sites, {\it infected core patches ({\it ICP})},  {\it infected peripheral patches ({\it IPP})} and {\it randomly distributed infected patches ({\it RDIP})}. 
		Our results show that infection spreads monotonically in the lattice with increasing migration without showing any sign of recovery in the {\it ICP} case. In the {\it IPP} case, it follows a similar monotonic progression with increasing migration, however, a self-organized healing process starts for higher migration, leading the lattice to full recovery at a critical rate of migration. Encouragingly, for the initial {\it RDIP} arrangement, chances of recovery are much higher with a lower rate of critical migration. An eigenvalue based semi-analytical study is made to determine the critical migration rate for realizing a stable infection-free lattice. 
		The initial fraction of infected patches and the force of infection play significant roles  in the self-organized recovery. They follow an exponential law, for the {\it RDIP} case, 
		that governs the recovery process. For the  frustrating case of {\it ICP} arrangement, we propose a random rewiring of links in the lattice allowing long-distance migratory paths that effectively initiate a recovery process. Global prevalence of infection thereby declines and progressively improves with the rewiring probability that follows a power law with the critical migration and leads to the birth of emergent infection-free networks.
		
	\end{abstract}
	
	\maketitle
	
	\section{\label{sec:level1} Introduction }
	
	Spatial spreading of infections has been studied using the concept of information flow in complex networks \cite{jesus2012, ijbc2012, barthelemy2005,markov2010, newman2002spread, shirley2005impacts, salathe2010dynamics, gomez2011nonperturbative, wang2016suppressing, demirel2017dynamics, soriano2018spreading}. It is a serious  issue of concern  for  life and society, and it is important to trace the disease spreading processes in human population  naturally occurring through air-traffic network \cite{colizza2006role,tatem2006global}, animal population network \cite{
		lentz2016disease,webb2006investigating}, insect population network \cite{stroeymeyt2014organisational,naug2008structure} and ecological network \cite{shaw2014networks,golhani2018review}. Usually, the network considered in disease modeling is very complex \cite{M06}.  Investigations of epidemics in  stochastic lattice model have been found in the literature with their elegant mathematical arguments and techniques which are used to understand the  critical threshold for the onset of epidemics \cite{grassberger1983critical, pastor2015epidemic, rhodes1997epidemic, keeling2005networks}.
	Asynchronous stochastic models have also been studied where all the sites are separated into susceptible and infected nodes and a Monte Carlo simulation has been performed by picking an infected site randomly, transforming it into susceptible one if it is below an infection threshold, otherwise assuming as infecting the randomly chosen nearest neighbors 
	\cite{tome2010critical, de2010stochastic}. These studies lead to an understanding of the spreading processes of disease in a large population over a long time. However, these models did not incorporate the deterministic variation of the susceptible-infected-recovered (SIR) or the susceptible-infected-susceptible (SIS) description of individual patches and unable to provide detail information on the impact of migration of population and especially, on the initial distribution pattern of infected  sites in a network.
	\par Alternatively, we consider here a deterministic  dynamical disease model  over the top of a square lattice, a simpler network where each node is interacting only with its nearest neighbors. We  consider an initial outbreak of infection in a fraction of sites in the lattice  with a particular pattern and  study the epidemic progression in the entire lattice through migration of population. Each site of the lattice is considered as a patch  and a disease spreads within a single patch following the SIS type deterministic model. The migration between the  connected patches is described by a diffusive dispersal process controlled by the rate of migration. Earlier studies \cite{hens2019spatiotemporal, brockmann2013hidden}  considered complex networks of patches to investigate disease spreading from a locally perturbed source, where infection starts from a single patch.  However, the possibility of infection in  multiple patches, at one time, in the  lattice and the role of migration were  not focused. 
	In our  proposed lattice model, nodes or patches are as usual connected with their immediate neighbors, and both the susceptible and infected hosts are allowed to move between the adjacent patches as controlled by their respective rates of migration. A disease-free patch/node, therefore, may be invaded by some parasites through migration of infected hosts between the immediate neighboring patches. 
	\par
\begin{figure}[!h]
		\centering	
		\includegraphic[height=3cm,width=3.15cm]{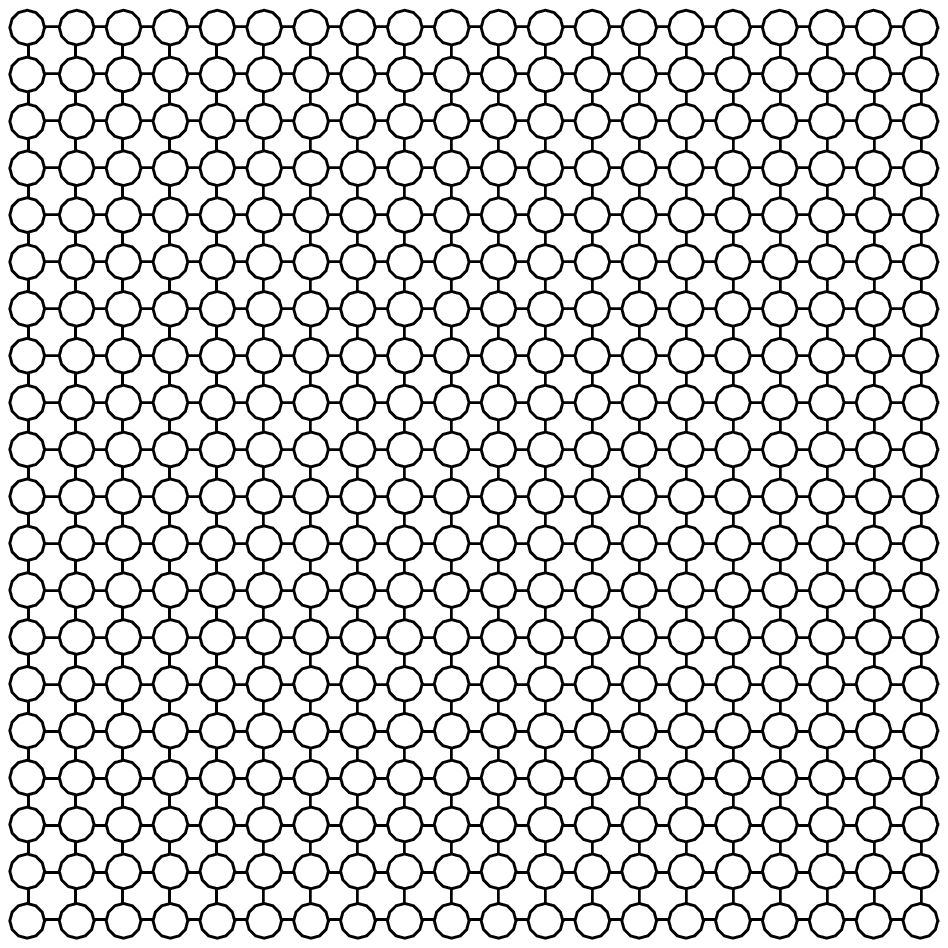}{(a)\; \quad infection-free}{0pt}{0.1}
		\includegraphic[height=3cm,width=3.15cm]{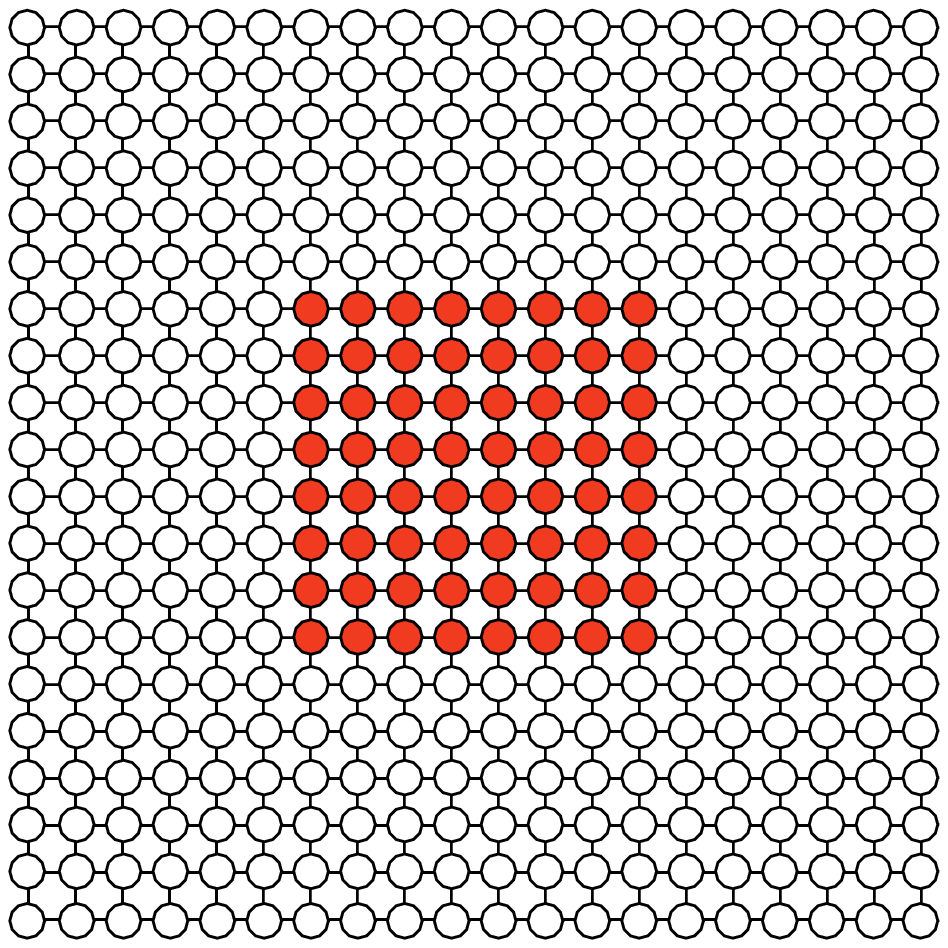}{(b) \;\quad\quad {\it ICP}}{0pt}{0.1} 
		\includegraphic[height=3cm,width=3.15cm]{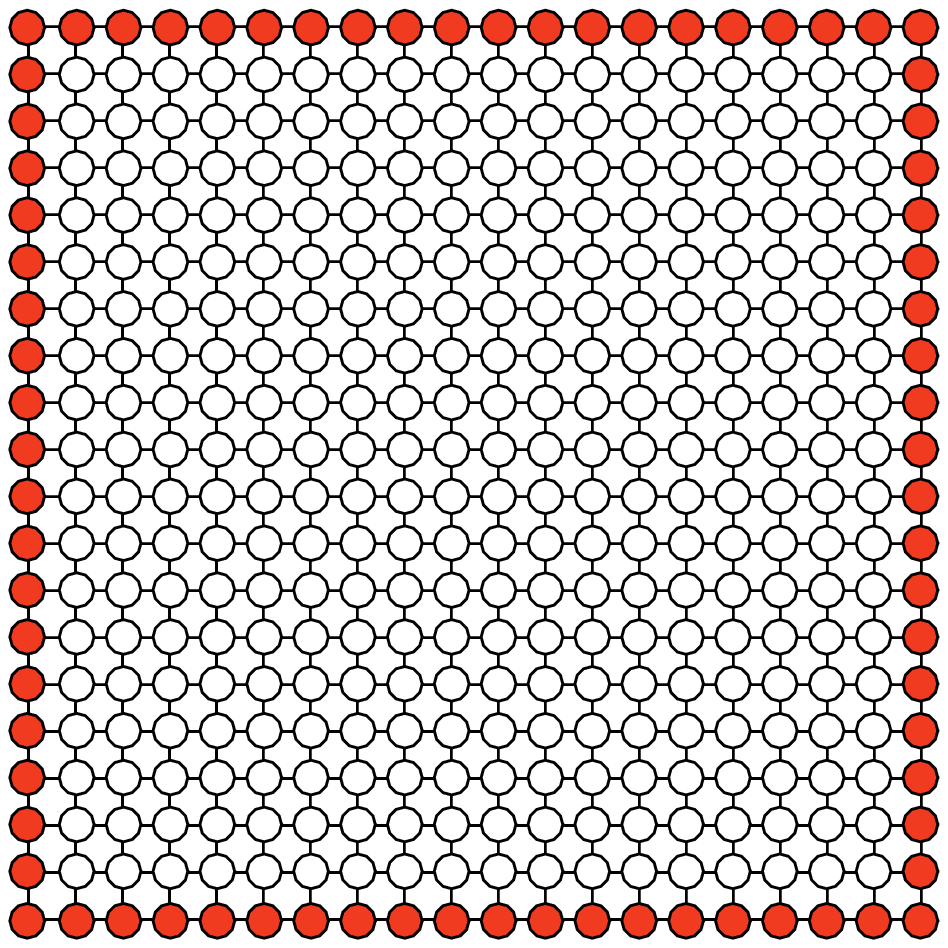}{(c) \;\quad\quad {\it IPP}}{0pt}{0.1}
		\includegraphic[height=3cm,width=3.15cm]{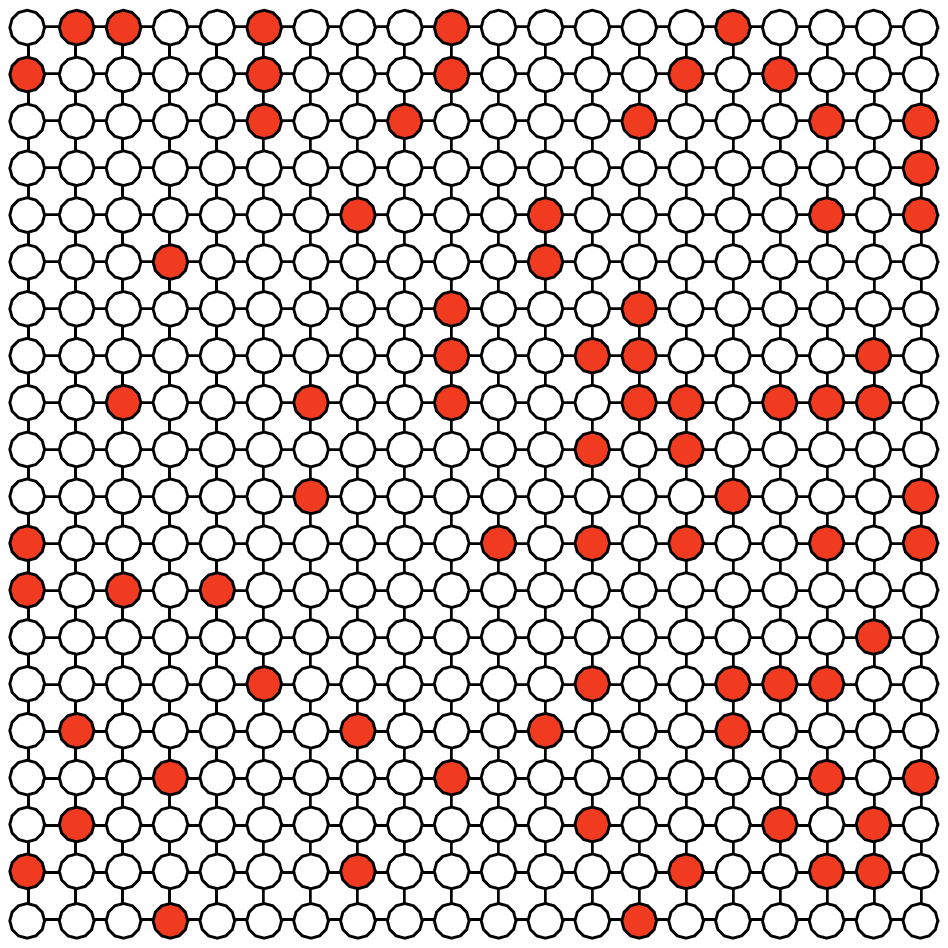}{(d) \; \quad\quad {\it RDIP}}{0pt}{0.1}
		\caption{Square lattice of patches. Examples of (a) infection-free, (b) infected core patches ({\it ICP}), (c) infected periphery patches ({\it IPP}), (d) randomly distributed infected patches ({\it RDIP}). White and red circles  exhibit sites in stable infection-free and infected  steady states, respectively. Each site or patch at the periphery has three neighbors except the four corner patches, having two neighbors each, and the rest of the patches have four neighbours.} \label{lat}
	\end{figure}

 Transmission of disease from one infected node to an infection-free  node and the severity as well as the global disease persistence largely depend upon population migration. The increased migration  has been shown to be beneficial for  an infected meta-population with an initiation of a recovery process \cite{GET02,MD02}. Experimental results \cite{LET05} in a metapopulation structure with flour beetle and its ecto-parasitic mite show that both the local and global prevalence of parasites depend on host migration rate and its local density. On the contrary, increased migration has also been shown to increase the host's extinction risk in host-pathogen meta-population model \cite{HET12}. However, it has not been explored whether the initial number of infected patches and their distribution pattern, in particular, can play any role in the spreading  and recovery of infection in a network. How do the rate of migration and the forces of infection jointly affect the local and global persistence of an infection in a square  lattice with a different initial distribution of  infected patches? Does there exist any correlation between the  forces of infection or migration rate and the number of initial infected patches in the eradication process? These questions are vital and need to be addressed appropriately for understanding a disease progression as well as  the implementation of any preventive measure. Our endeavour here is to address such pertinent questions where the disease spreads in a square lattice with either of the three presumed sets of initially arranged distribution of infected sites: (i) infected sites at the center called as {\it infected core patches} ({\it ICP}), (ii) infected sites at the periphery called as {\it infected-periphery patches} ({\it IPP}) and (iii) infected sites randomly distributed in the lattice or {\it randomly distributed infected patches} ({\it RDIP}). Exemplary initial  distribution patterns of infected patches in the lattice are shown in Fig.~\ref{lat}.
	\par Our study shows nontrivial results of disease spreading and recovery that predominantly depends on the initial arrangement of infected patches. For the start of disease from centrally located sites ({\it ICP} case), the infection spreads monotonically in the lattice without any sign of recovery with increasing migration. For an initial infection in the peripheral sites ({\it IPP Case}), disease spreading also increases monotonically, however, a healing process starts with a higher rate migration and a full recovery is possible at a larger critical rate of migration. When the infection starts in random locations ({\it RDIP} case), spreading is faster and infection covers most of the sites at a lower rate of migration, however, the lattice also recovers at a lower critical rate of migration. Furthermore, we have checked the role of the force of infection or transmission efficiency, for all three initial distributions, since it is one of the most important parameters in epidemic models \cite{HET10}. This epidemiological trait describes how rapidly the infection spreads within a patch and thereby increases the burden of disease. We explore the combined effect of the rate of migration and the force of infection for all three arranged set of initially infected patches, separately. 
	Persistence and non-persistence of infection in the lattice also depend on the number of initially infected patches, which we have given attention in our study. For the {\it RDIP} distribution of initial infection, in particular, the number of initially infected patches of the lattice for which it becomes infection-free and the relative infectivity of disease follows an exponential law.
	\par  Thus {\it RDIP} distribution of initial infections shows the highest recovery efficiency and the {\it ICP} has a worse record. The initial infection in centrally located sited  {\it ICP case} is simply incompetent to remove infection by a natural or a self-organized  process  except for very low transmission efficiency or force of infection. To circumvent this situation, we suggest a random process of rewiring the links \cite{sw} in the lattice and it shows a promising result for the case of initial {\it ICP} arrangement  of infected sites; it is eventually able to remove the infection with an emergent network at a critical migration rate, which follows a power law with the rewiring probability. 
	It is to be cautioned that for a significantly high force of infection, none of the initial arrangements of infected patches can deter the progress of the disease or help start a recovery process. 
	\par We elaborate the lattice model in Sec.I I. The main results on the impact of three initial distributions of infected sites are presented in Sec. IIIA, the combined role of migration and force of infection in Sec. IIIB, discussed the role of the initial fraction of infected sites in Sec. IIIC. The rewiring probability and the infection-free emergent network have been elaborated in Sec. IIID. The conclusive statements are made in Sec. IV. An Appendix is added at the end to discuss the impact of a threshold of initial density of infected population, time evolution of all the hosts in a long time, in all three cases, and the eigenvalue analysis to determine the critical migration rate for realizing a stable infection-free state.
	
	\section{Susceptible-Infected-susceptible model in a square lattice}	
	We consider $N \times N$ patches or sites in a square lattice, where both the susceptible and infected hosts are allowed to move from one patch to its  neighbouring patches of the lattice. The  dynamics of susceptible host, $x_i$, and infected host, $y_i$, in the $i^{th}$ patch $(i=1,....,N \times N)$ and their dispersal in the lattice are described by
	\begin{align}
	\frac{{d{x_i}}}{{dt}} = & {b}\left( {{x_i} + {y_i}} \right)\left( {1 - \frac{{{x_i} + {y_i}}}{{{c}}}} \right) - {d}{x_i} + \gamma {y_i} - {\beta }{x_i}{y_i} \nonumber \\
	& + {\epsilon _1}\sum\nolimits_{j = 1}^{N{\times}N} {{A_{ij}}({x_j} - {x_i})},  \nonumber \\
	\frac{{d{y_i}}}{{dt}} = &{\beta }{x_i}{y_i} - \left( {{d} + \gamma  + \alpha } \right){y_i} + {\epsilon _2}\sum\nolimits_{j = 1}^{N{\times}N} {{A_{ij}}({y_j} - {y_i})} \label{eq2}, 
	\end{align}
	where  $A_{ij}$ is the adjacency matrix that defines the network structure, and $\epsilon_1$ and $\epsilon_2$ are the  diffusion or migration rates of susceptible and infected hosts, respectively. The local dynamics of each isolated patch (when $\epsilon_1=0=\epsilon_2$) is governed  by the SIS type epidemic model, where the hosts follow density-dependent regulation with $b$ as the growth rate  and $c$ as the environmental carrying capacity.  A micro-parasitic infection, having disease transmissibility $\beta$ and recovery by a natural process (host defense mechanisms) at a rate $\gamma$, spreads through contacts between infected and susceptible hosts with no vertical transmission. The parameters $d$ and $ \alpha$ are the natural death rate of host and virulence of the disease, respectively. It is to be mentioned that individuals who leave one patch immediately enter the other patch, thereby avoiding population death during dispersal. The dispersal and vital dynamics have the same time scale so that dispersal can rescue populations on individual patches from infection and at the same time dispersal can also make an infection-free patch infected. Open boundary condition is considered for the lattice. All parameters are non-negative. The single patch model was earlier analyzed from the evolutionary point of view  \cite{vB98} and later extended to a two-patch model using variable aggregation method (slow-fast dynamics) \cite{charles2002host}.
	
\begin{figure}[H]
		\centering	
		\includegraphic[height=4cm,width=5cm]{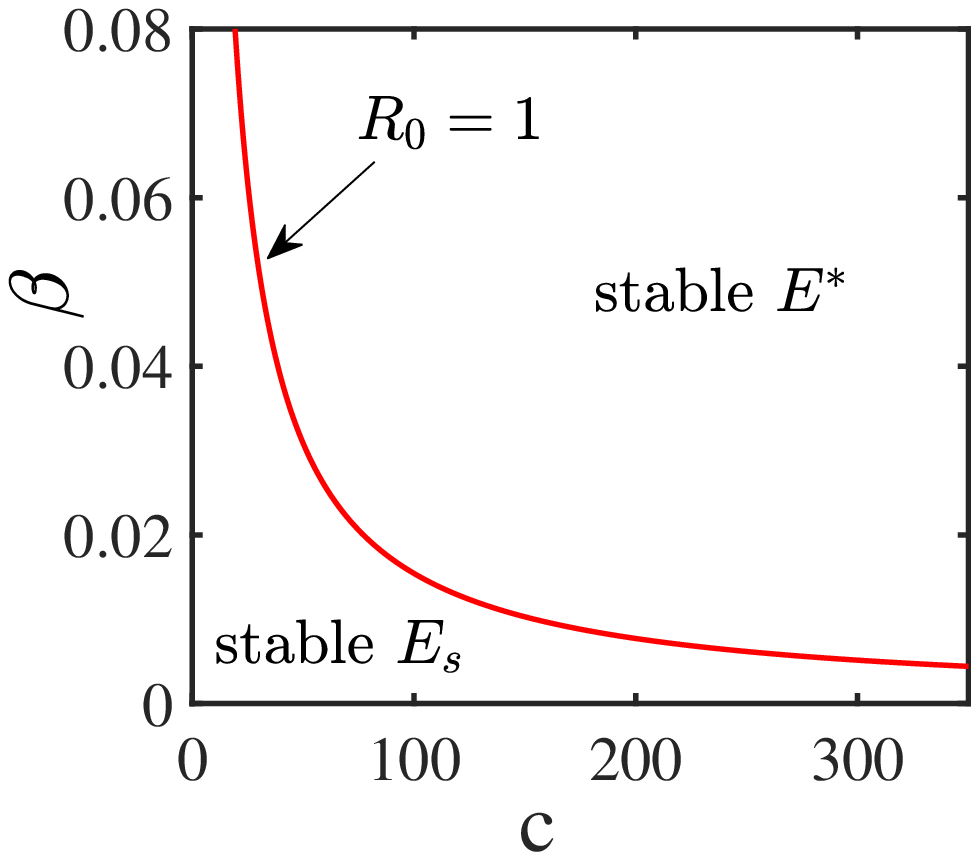}{(a)}{125pt}{1.3}
		\includegraphic[height=4cm,width=6cm]{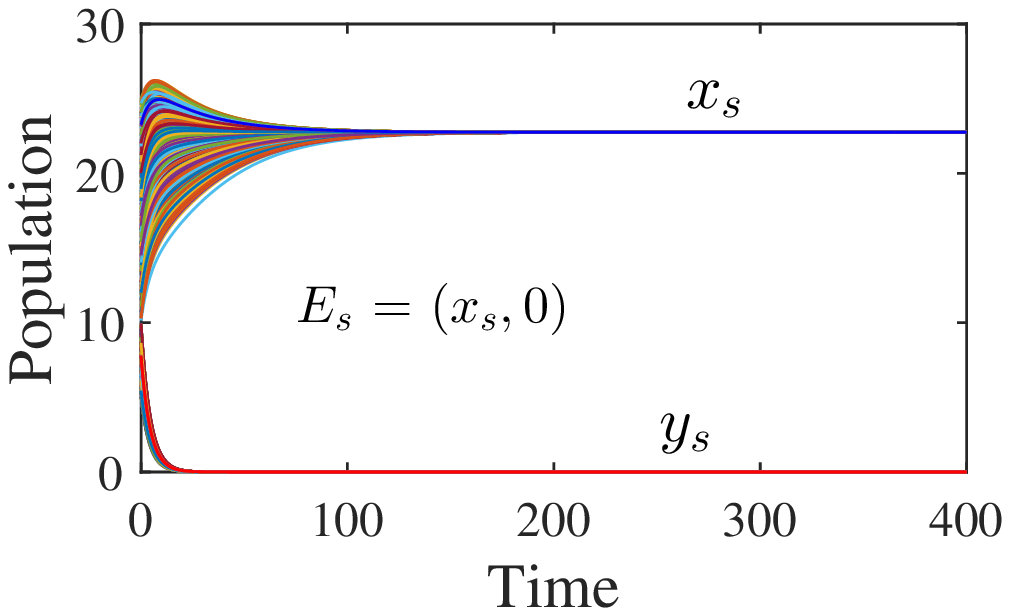}{(b)}{145pt}{1.3}
		\includegraphic[height=4cm,width=6cm]{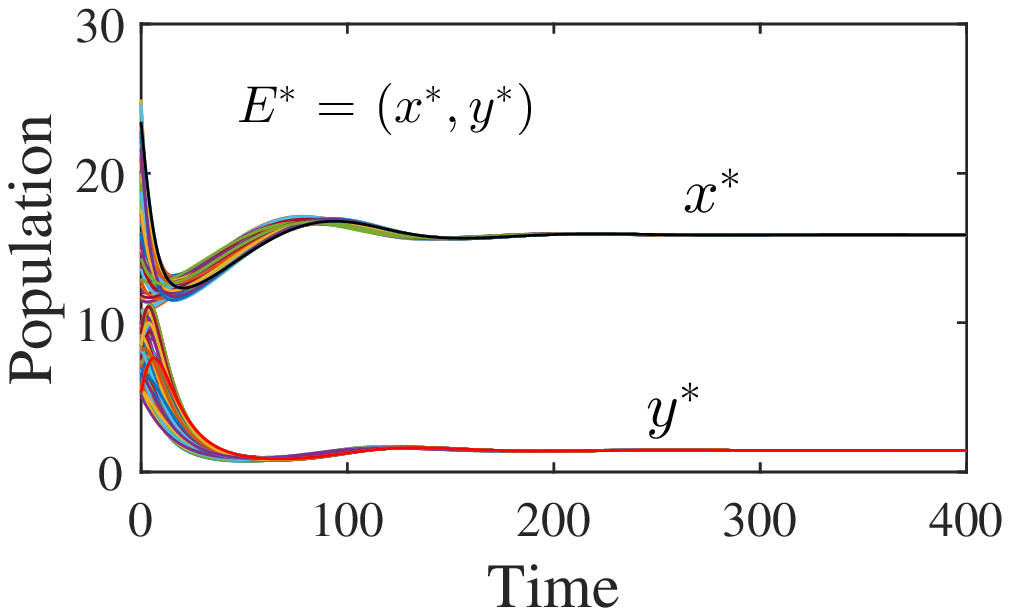}{(c)}{145pt}{1.3}
		\caption{(a) Bifurcation of a single patch model in $c-\beta$ plane. Stability regions of two steady states $E_s$ and $E^*$ are separated by the transcritical bifurcation line $R_0=1$.   (b,c) Temporal evolution of two subgroups of patches when $\epsilon_1=0=\epsilon_2$.  (b)  One subgroup stabilizes to infection-free equilibrium $E_s(x_s,0)$  for $\beta_{0_m}=0.002<\beta_c$, where $\beta_c=0.0119$, (c) second subgroup of the remaining patches stabilizes to endemic equilibrium $E^*(x^*, y^*)$  corresponding to the condition, $\beta_{inf_n}=0.017>\beta_c$. Parameters are $b=0.2, c=130, d=0.165, \alpha=0.01, \gamma=0.005$. Initial conditions are $x(0)_i\in [10,25]$ and $y(0)_i\in [1,5]$.}  \label{fig2}
	\end{figure}

	The study is initiated by dividing the entire lattice into two subgroups, viz. infection-free and infected patches by choices of the force of infection, $\beta$. Such choices of $\beta$ can be made from the dynamics of single patch model, which has three equilibrium points. In an isolated patch, an equilibrium $E_0(0,0)$ always exists and unstable for all feasible parameters. The birth rate shall always exceed the death rate of hosts (i.e., $b>d$) for the existence of any positive equilibrium; an infection-free equilibrium  $E_s(x_s,0)$  exists, where $x_s=\frac{c(b-d)}{b}$. The parasites are unsuccessful to invade the host population (i.e., $E_s$ is stable) if $R_0<1$, where $R_0=\frac{\beta c(b-d)}{b(d+\gamma+\alpha)}$ is the basic reproduction number, measuring the average number of secondary infection produced by a single infected host during its infectious period when introduced into a group of susceptible population \cite{AM01}. The equivalent condition in terms of force of infection can be expressed by $\beta<\beta_c$, where $\beta_c=\frac{b(d+\gamma+\alpha)}{c(b-d)}$. The equilibrium host densities of an endemic equilibrium point $E^*(x^*,y^*)$ are
	$x^*=\frac{d + \gamma+ \alpha }{\beta}$ and $x^*+y^*=\frac{c}{b}[(b-d)-\frac{b\alpha y^*}{x^*+y^*}]<x_s$. The last inequality implies that the total equilibrium host population is suppressed in the presence of infection compared to infection-free state equilibrium value, $x_s$.
	Thus the dynamics of a single patch SIS model is entirely determined by the basic reproduction number, $R_0$. In particular, if $R_0>1$ (or equivalently, $\beta>\beta_c$) then infection persists and the endemic equilibrium becomes stable whereas disease cannot persist in the opposite case (see Fig.~\ref{fig2}(a)).
	Therefore, the force of infection is considered as $\beta={\beta_0}_m<\beta_c$ for the subgroup of infection-free patches of the lattice, and that for the subgroup of infected patches, it is considered as $\beta=\beta_{inf_n}>\beta_c$ ($m,n$ are pseudo-indices for parameter $\beta$ of infection-free and infected patches, respectively).
	The condition $\beta_{0_m}<\beta_c<\beta_{inf_n}$ differentiates patches into two states, viz. an infection-free state (Fig.~\ref{fig2}(b)) and an infected state (Fig.~\ref{fig2}(c)) under no migration ($\epsilon_1=0=\epsilon_2$). Thus, patches are characterized by the disease transmissibility parameter $\beta$ and not by the initial value of the infected population. An infection-free patch is, therefore, a patch with $\beta<\beta_c$ and an infected patch is a patch where $\beta>\beta_c$.  
	
	\section{Infection spreading and Recovery: Numerical Results}
\begin{figure*}[!h]
		\centering	
		\includegraphic[height=2.5cm,width=2.5cm]{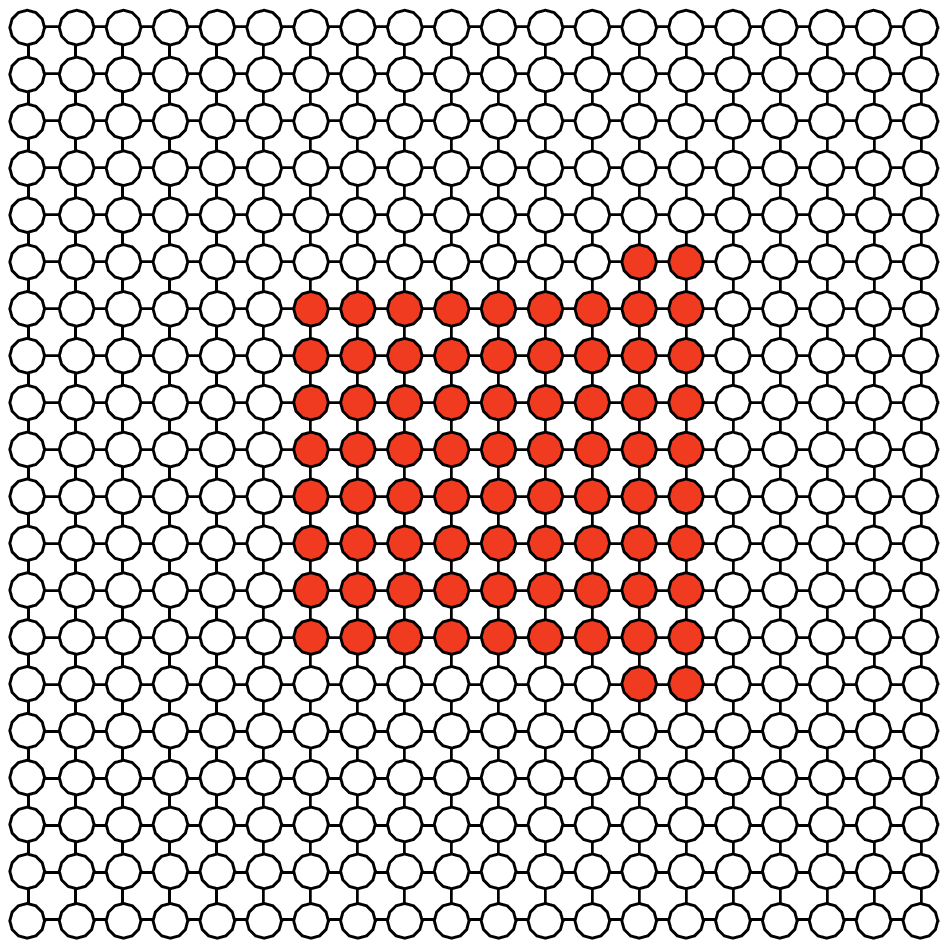}{(A1)  $\epsilon=0$}{0pt}{0}
		\includegraphic[height=2.5cm,width=2.5cm]{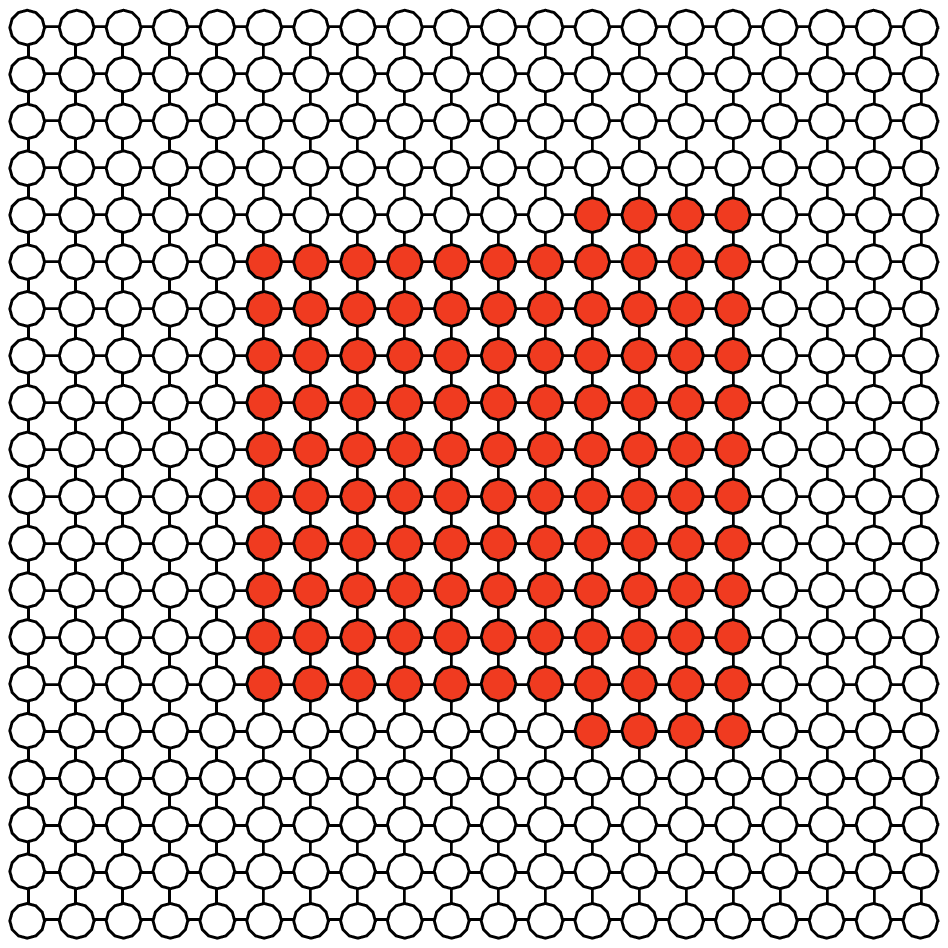}{(A2) $\epsilon=0.05$}{0pt}{0}
		\includegraphic[height=2.5cm,width=2.5cm]{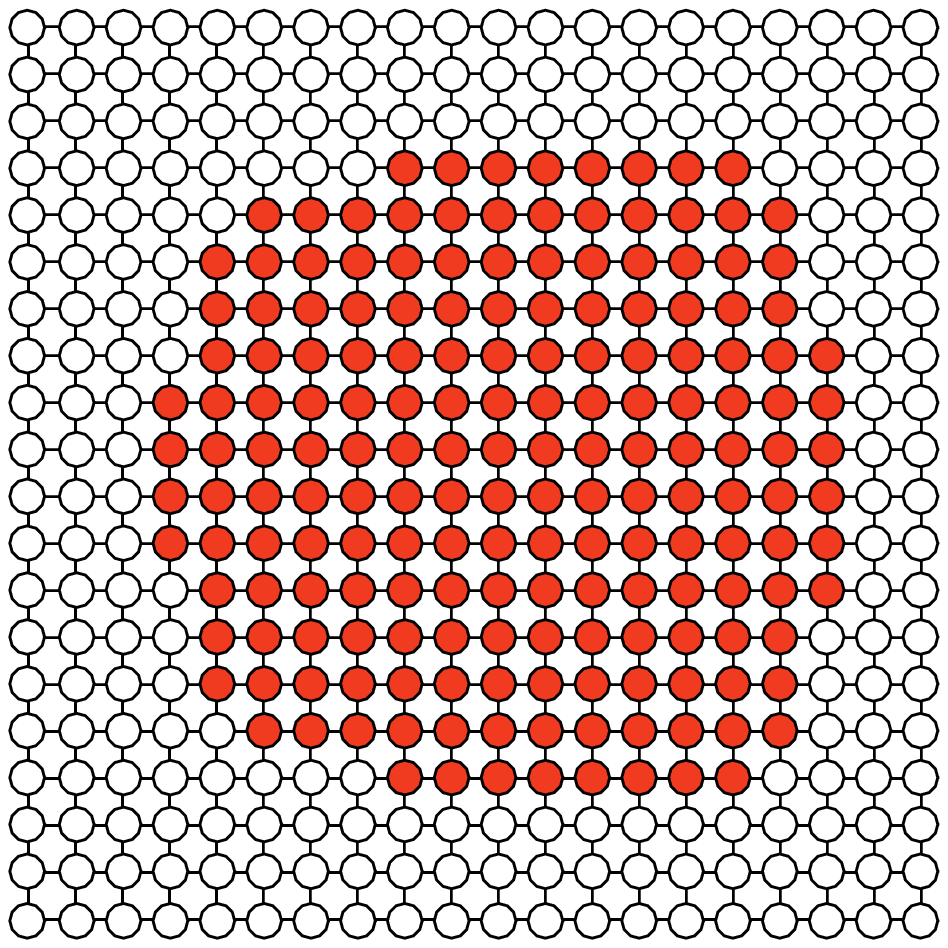}{(A3)  $\epsilon=0.2$}{0pt}{0}
		\includegraphic[height=2.5cm,width=2.5cm]{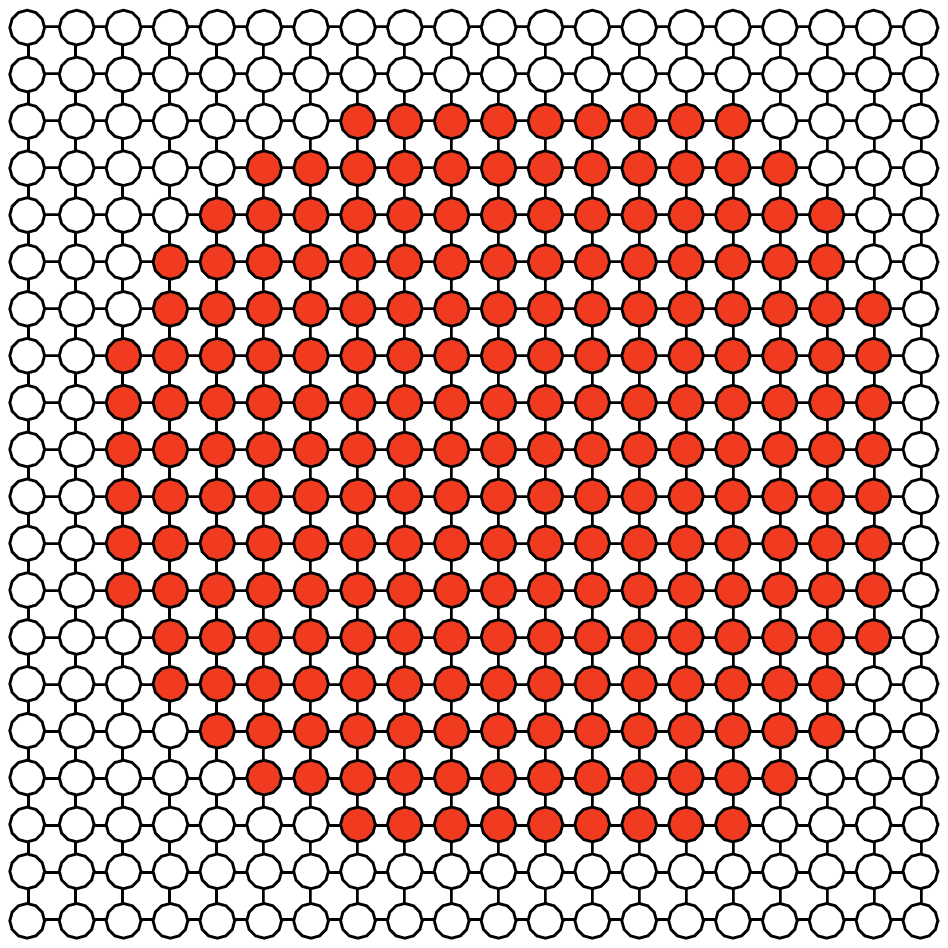}{(A4)  $\epsilon=0.5$}{0pt}{0}
		\includegraphic[height=2.5cm,width=2.5cm]{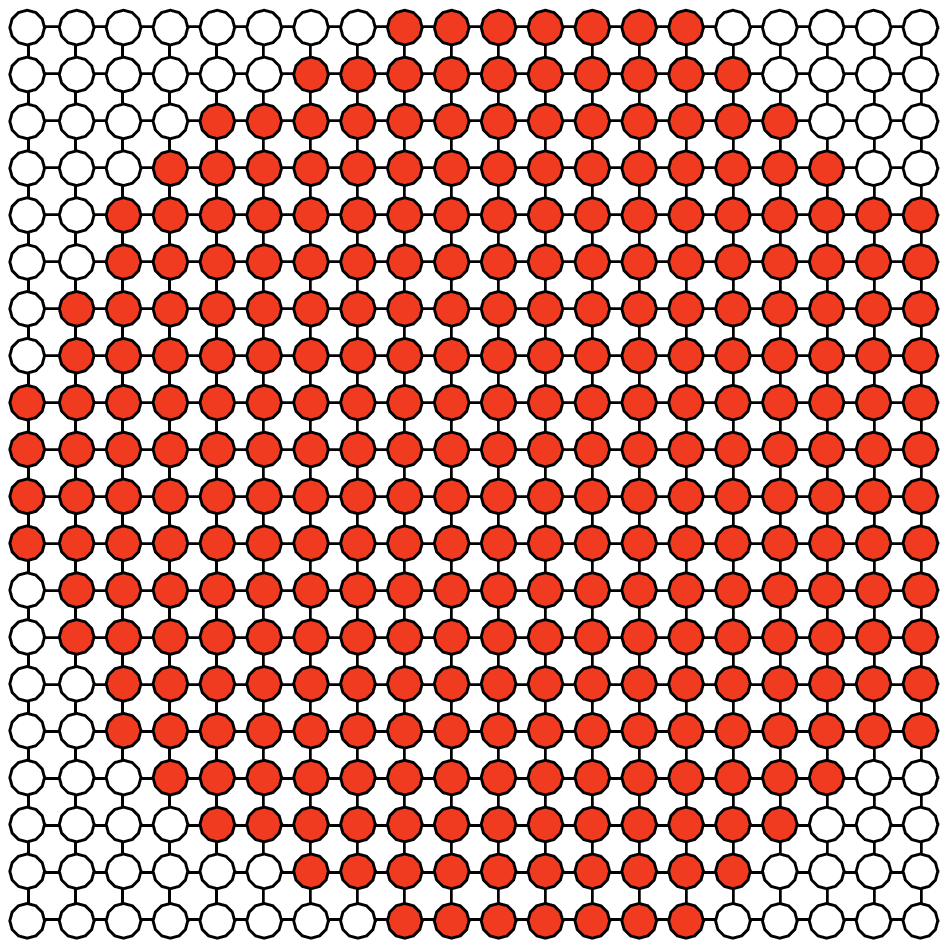}{(A5) $\epsilon=1$}{0pt}{0}\\
		
		\vspace{5pt}
		
		\includegraphic[height=2.5cm,width=2.5cm]{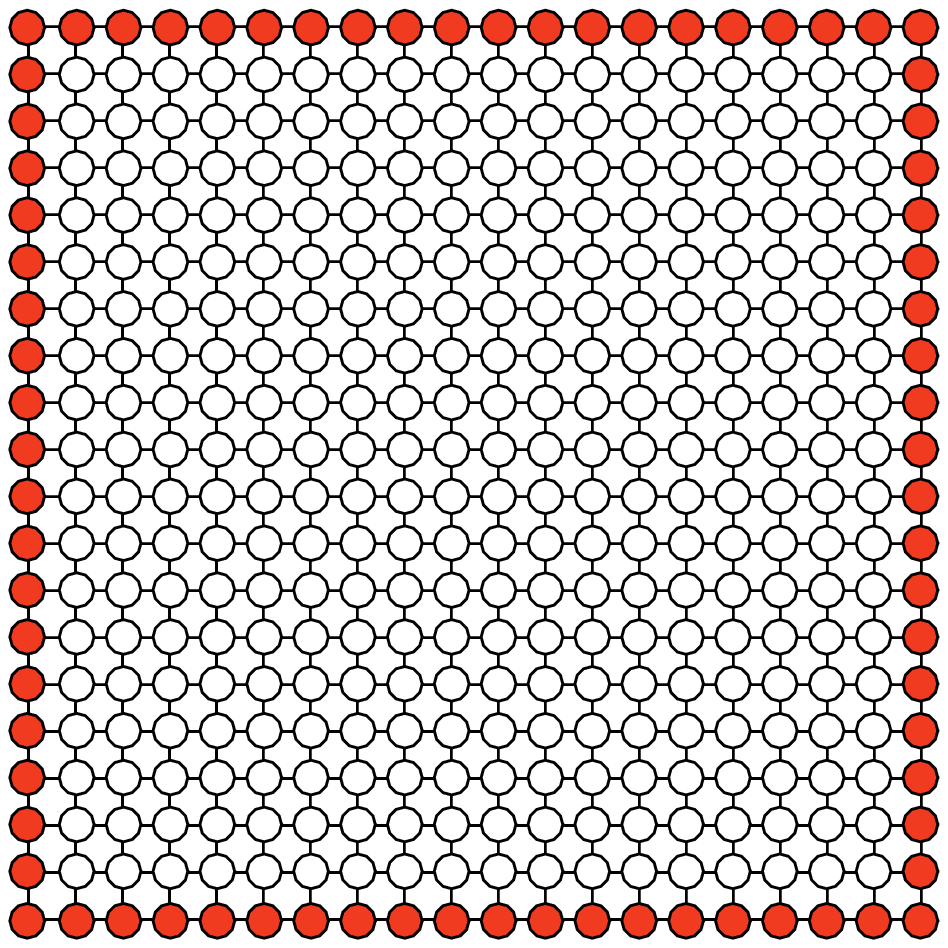}{(B1) $\epsilon=0$}{0pt}{0}
		\includegraphic[height=2.5cm,width=2.5cm]{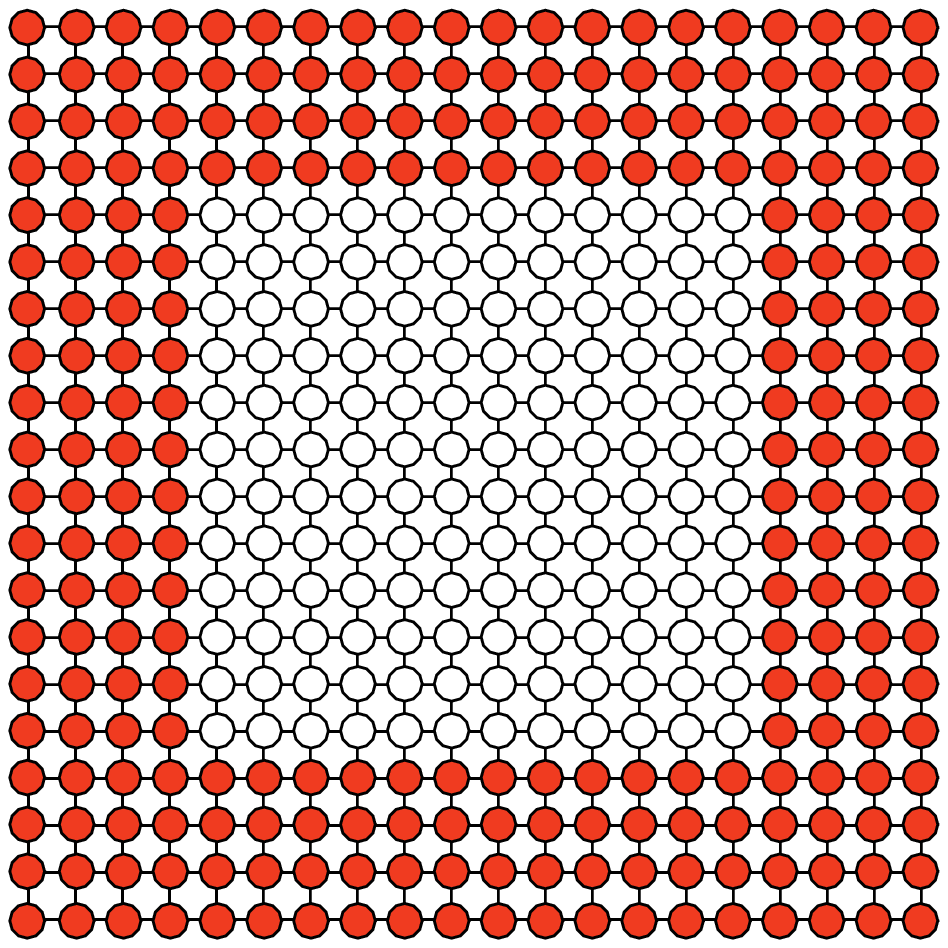}{(B2) $\epsilon=0.2$}{0pt}{0}
		\includegraphic[height=2.5cm,width=2.5cm]{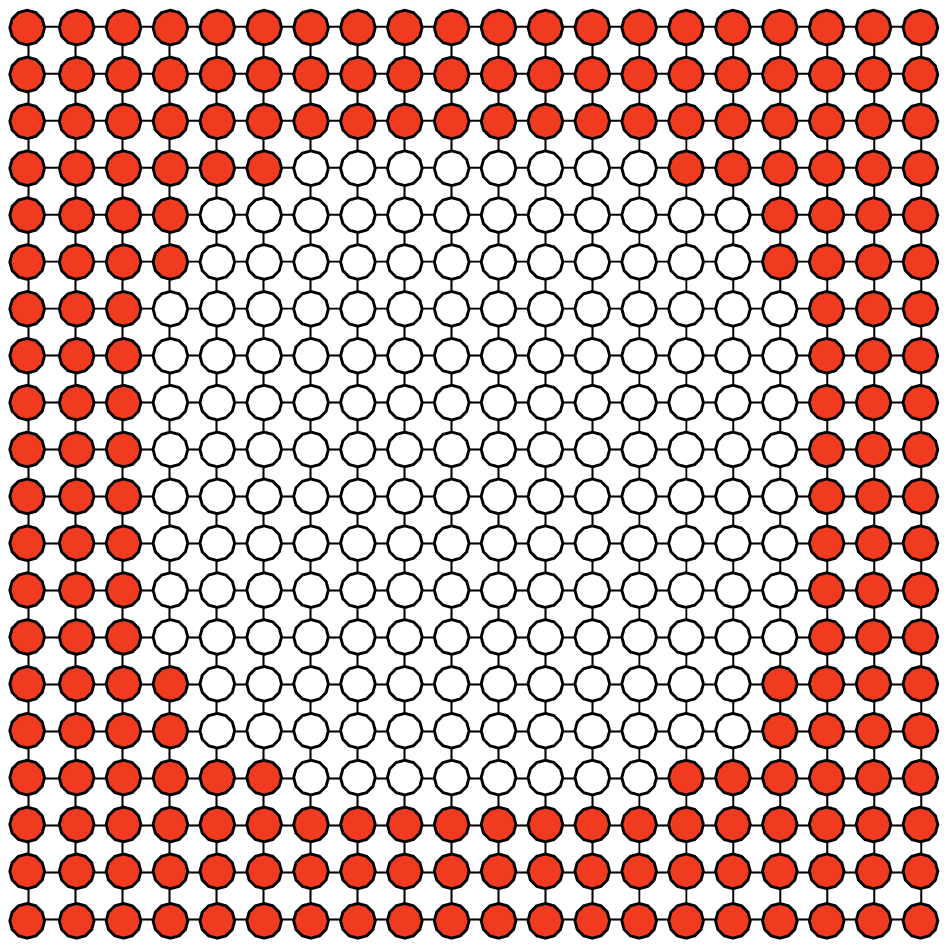}{(B3) $\epsilon=0.5$}{0pt}{0}
		\includegraphic[height=2.5cm,width=2.5cm]{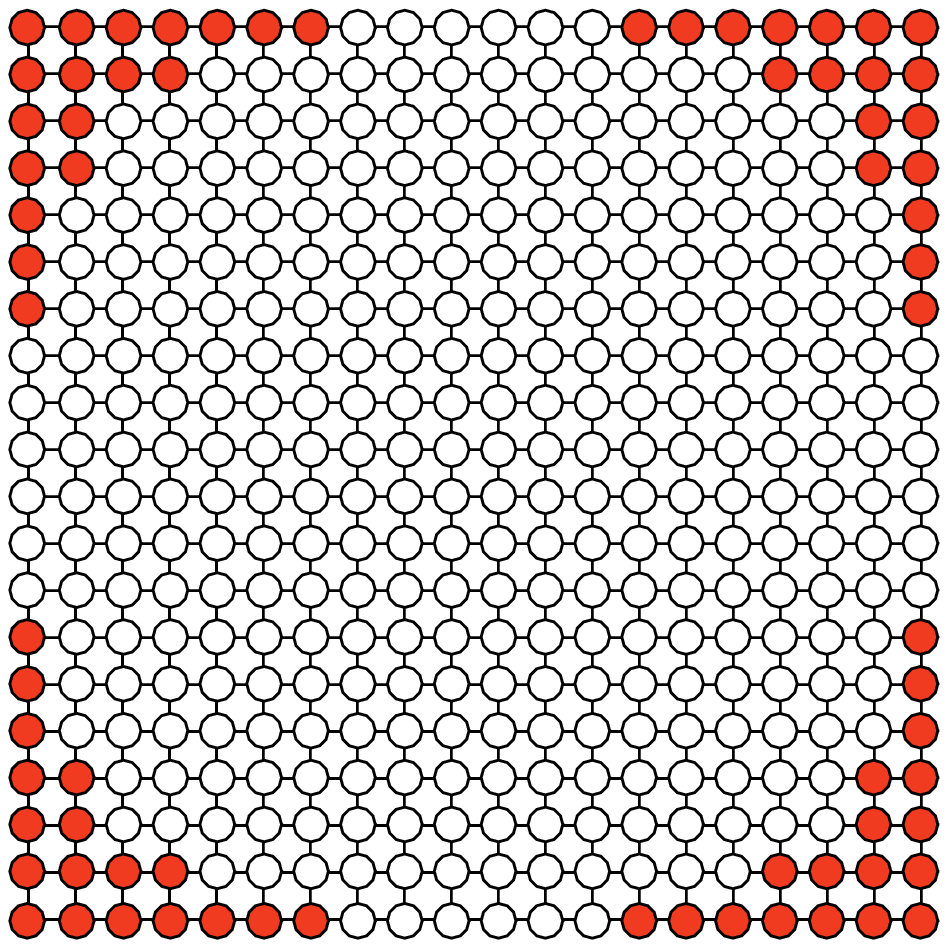}{(B4) $\epsilon=0.63$}{0pt}{0}
		\includegraphic[height=2.5cm,width=2.5cm]{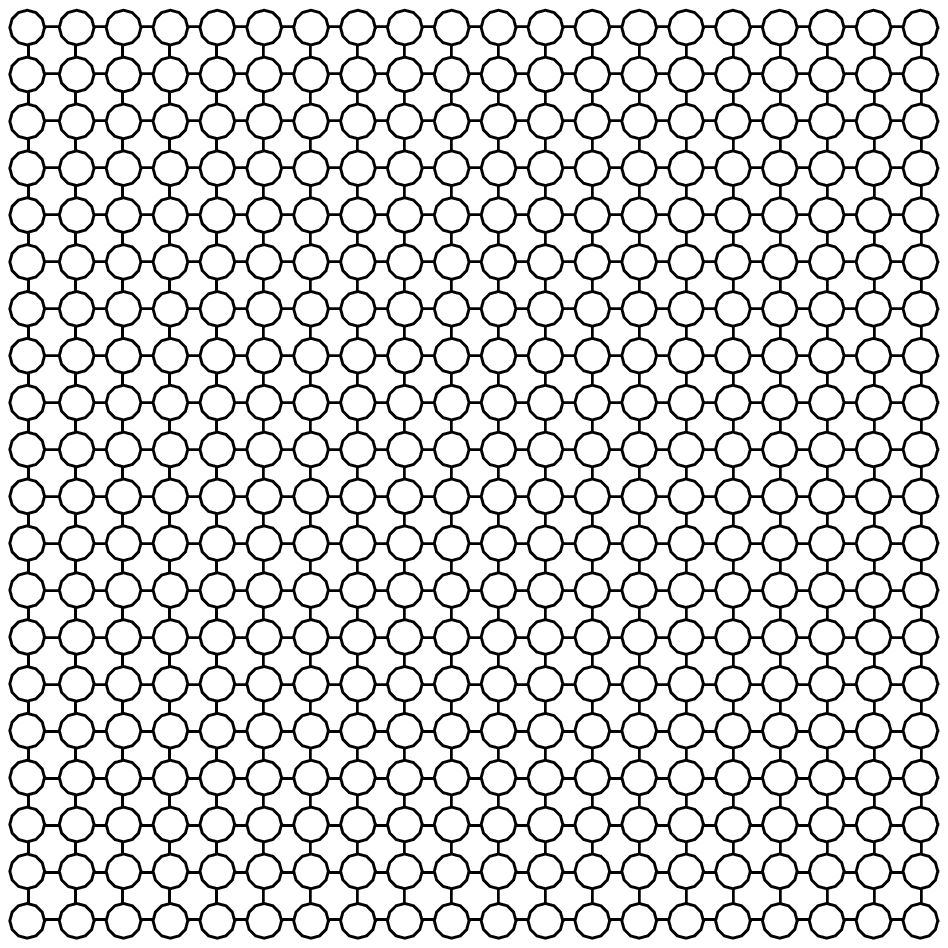}{(B5) $\epsilon=0.65$}{0pt}{0}\\
		\vspace{5pt}
		\includegraphic[height=2.5cm,width=2.5cm]{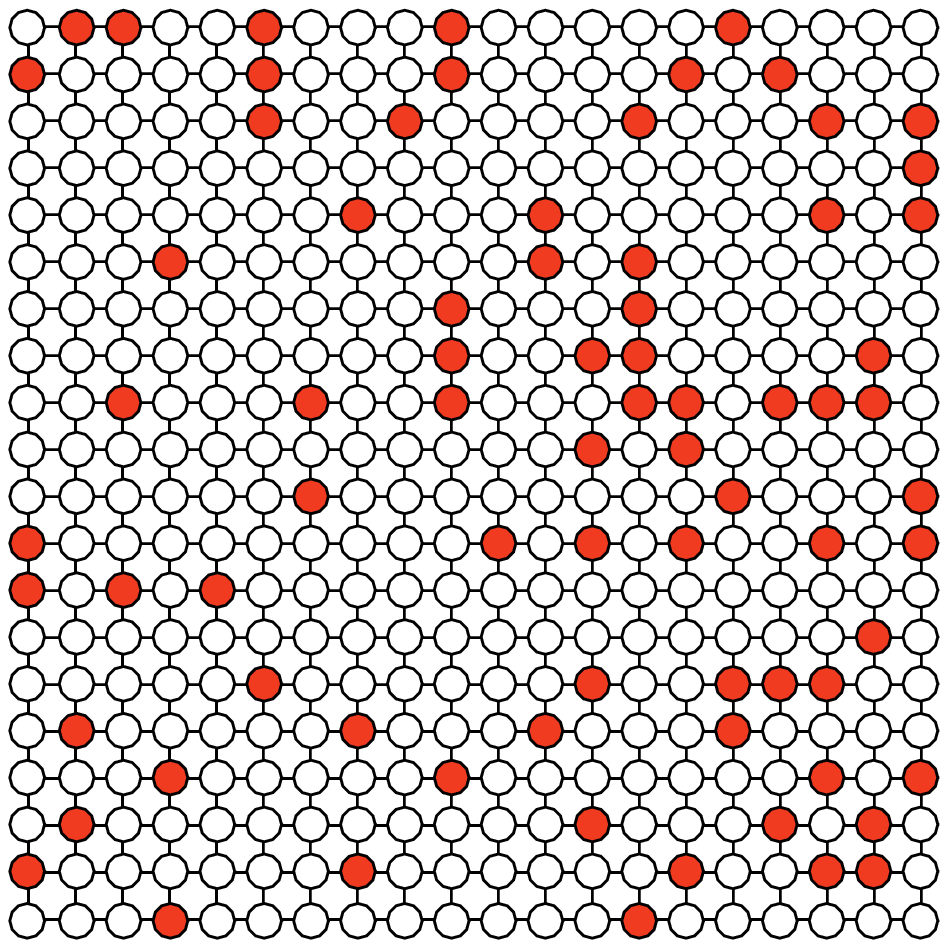}{(C1) $\epsilon=0$}{0pt}{0}
		\includegraphic[height=2.5cm,width=2.5cm]{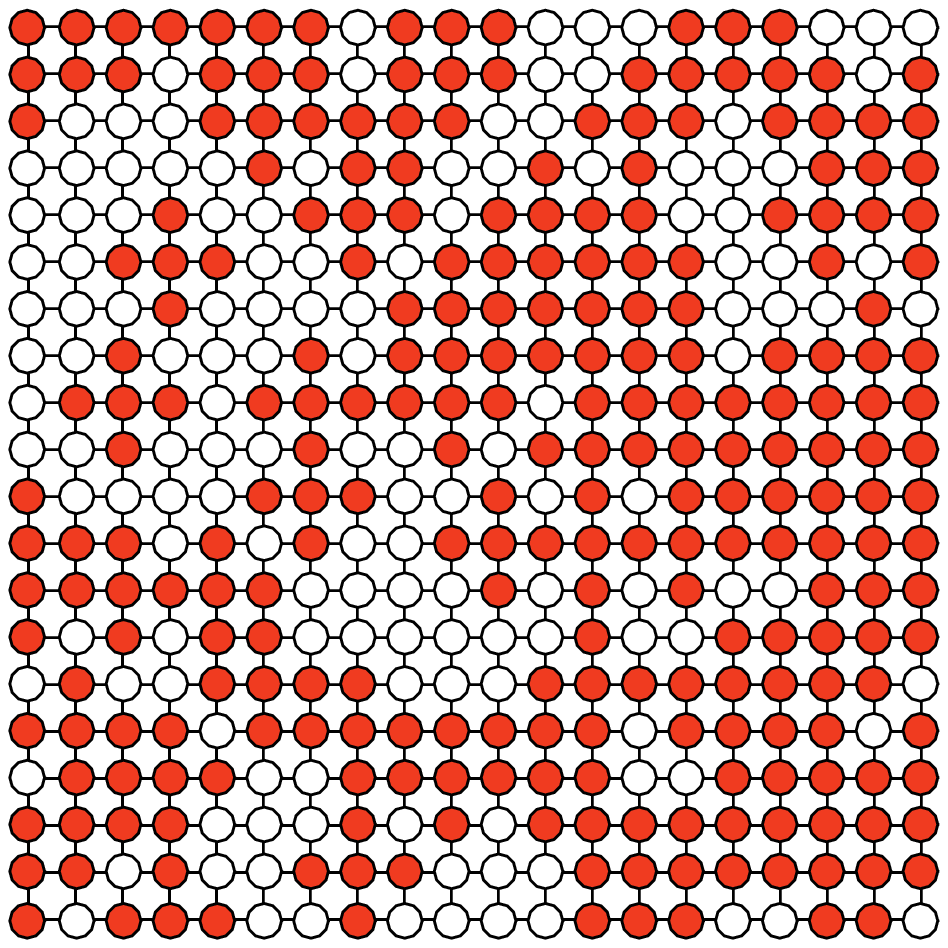}{(C2) $\epsilon=0.01$}{0pt}{0}
		\includegraphic[height=2.5cm,width=2.5cm]{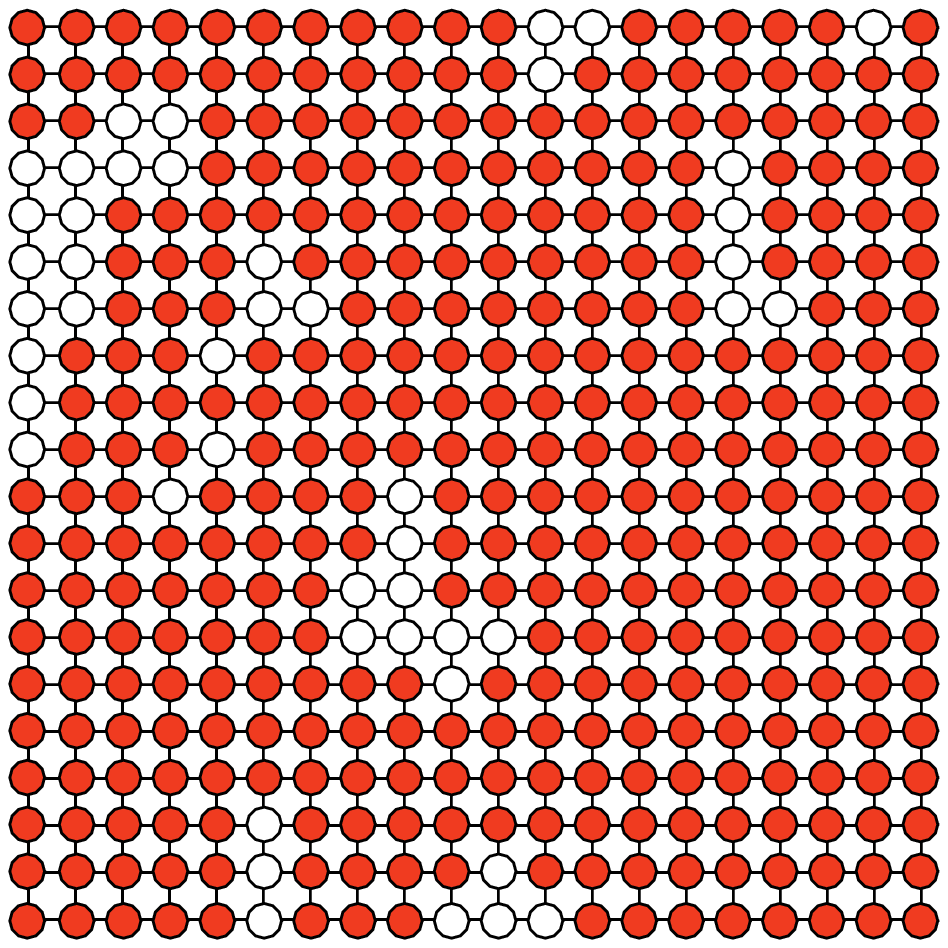}{(C3) $\epsilon=0.04$}{0pt}{0}
		\includegraphic[height=2.5cm,width=2.5cm]{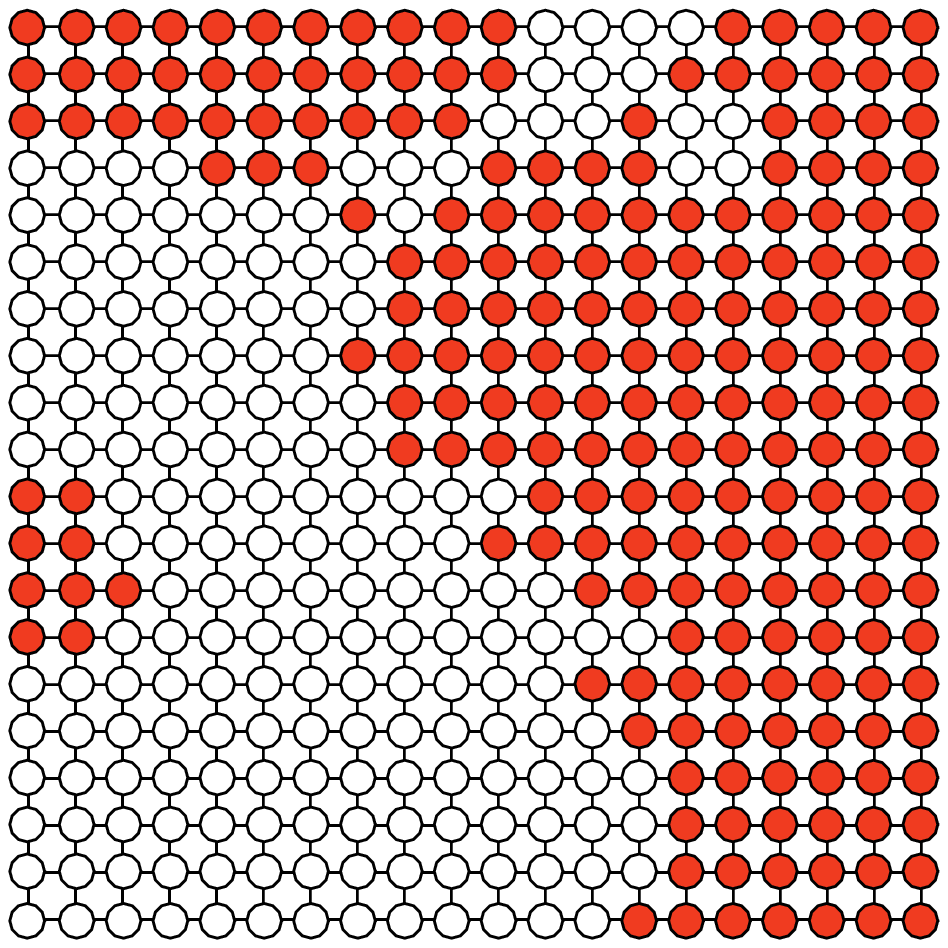}{(C4) $\epsilon=0.2$}{0pt}{0}
		\includegraphic[height=2.5cm,width=2.5cm]{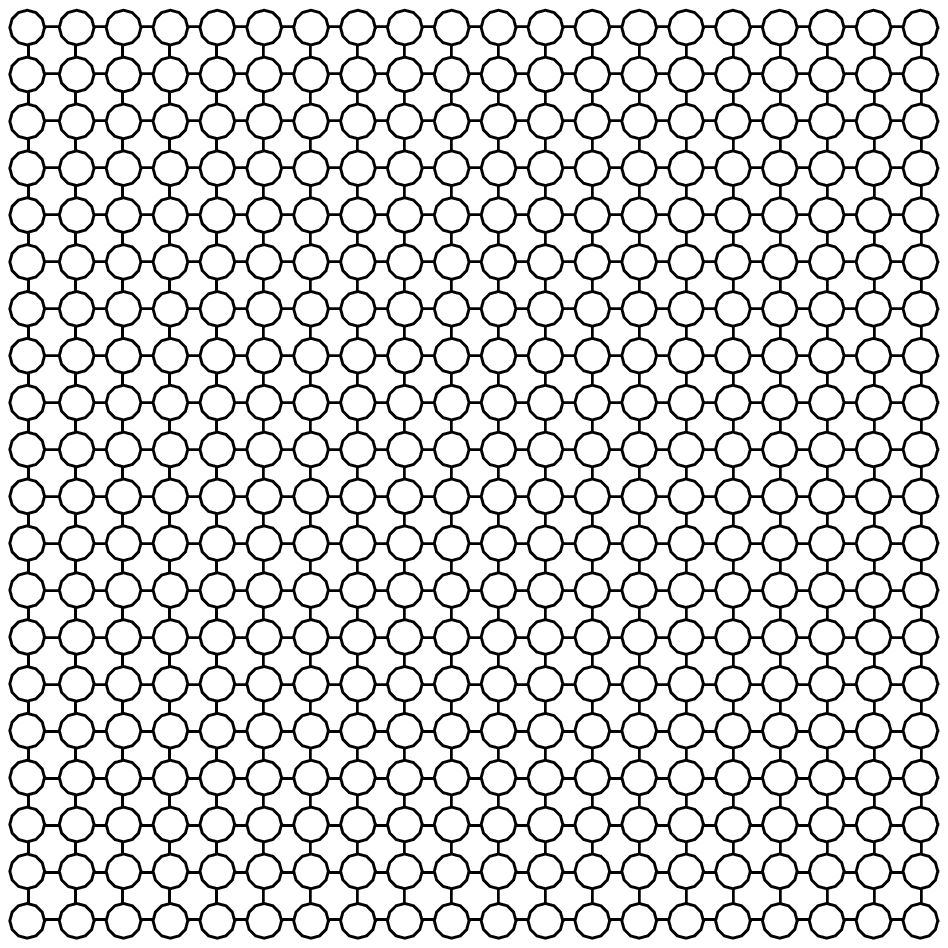}{(C5) $\epsilon=0.316$}{0pt}{0}
		\caption{Evolution of infected patches in a lattice with increasing migration rate $\epsilon$ $(\epsilon_1=\epsilon_2=\epsilon)$ for different initial  arrangements of infected patches. Upper row: Status of infection with increasing migration rate, when infection starts spreading from the core of the lattice ({\it ICP} arrangement). Middle row: Infection spreads from the periphery of the lattice ({\it IPP} arrangement).
			Lower row: Infection spreads from randomly distributed patches of the lattice ({\it RDIP} distribution). The number of initial infected patches is considered as $19\%$ of the total patches in all three cases. Force of infection is considered as $\beta_{0_m}=0.002$ for infection-free (white circles) patches and $\beta_{inf_n}=0.017$ in case of infected patches (red circles). Other parameters are as given for Fig.~\ref{fig2}.} \label{snap_grid}
	\end{figure*}

	A global scenario of infection spreading in a square lattice of $20\times20$ patches is presented for three proposed initial spatial arrangements of infections. 
	For simplicity, the migration rates of both the susceptible and infected hosts are considered as identical, $\epsilon_1$= $\epsilon_2$= $\epsilon$.  Our primary focus is on how initially arranged  pattern of infected patches affect the disease spreading and  recovery processes in the lattice under a migration of both the hosts. The rate of migration, the force of infection and their combined effects 
	are given prime importance, in this context, besides counting the influence of the initial number of infected patches. We considered only the asymptotic dynamics of the system (in our case, it is either an infection-free steady-state or an infected steady state) by discarding the transient state. A patch is colored red (see Fig.~\ref{snap_grid}) and said to be infected  if the final infected host density ($y_i$) remains equal or above a threshold, $y_h=0.01$, as determined by the asymptotic fixed point value of the patch. Otherwise, a patch is identified as infection-free, marked by white color, where the final value of infected hosts remains below the threshold, i.e., $y_i< y_h = 0.01$. The choice of the threshold ($y_h=0.01$) is made arbitrarily for numerical benefits. This restriction, however, can be relaxed and does not affect the main results as shown in Figs.~\ref{all2}(a-c) for a different threshold, $y_h=0.001$ (see Appendix A).
	We must note here that a few terminologies are exchangeably used, in the text, with the same meaning such as  patches as sites or nodes, force of infection or transmission efficiency or transmissibility.
	\subsection{Infection spreading under initial arrangements of infected sites}
	Our numerical results reveal  the spreading of infection in the lattice with varying  migration $\epsilon$ for our three  choices of initial arrangements of infected patches. Each numerical simulation (Fig.~\ref{snap_grid}) is initiated with $19\%$ infected patches (red color) and the rest of the patches are infection-free (white color), which are  defined by setting the parametric conditions, ${{\beta_{inf}}_n}>\beta_c$ and ${{\beta_{0_m}}}<\beta_c$, respectively.  
	Upper panels (A1)-(A5) of Fig.~\ref{snap_grid} demonstrate how the infection spreads from the core of the lattice with a gradual increase of migration. The number of infected patches grows monotonically as the migration rate increases and almost $75\%$ of the patches become infected at $\epsilon=1$. It indicates two important issues of concern. An outbreak of a highly infectious disease may turn into an epidemic if the outbreak occurs from centrally located patches of the lattice ({\it ICP} arrangement). The lattice never recovers  from the epidemic and no self-organized recovery process is seen to be initiated with migration. A disease control strategy, in such a case, needs to be adopted that shall include restricted migration and appropriate contact reducing technique or social-distancing of individuals. The raccoon rabies epidemic spreading in south-central state of the United States during the mid-seventies of the last century \cite{NET79} seems similar to the disease spreading process in the lattice with the {\it ICP} arrangement. It was reported that rabbis in the counties of the Kentucky state spread through rabies raccoon imported by private hunting clubs during 1975-1976. The disease subsequently spread over West Virginia, Virginia, Pennsylvania and other neighbouring states through  infected  migrants  and became the largest and most devastating wildlife rabies epidemic \cite{CET00}. However, it was endemic and confined to Florida, a peripheral province of USA,  for a long time since 1947 and spread towards the north and the central provinces during 1960-1970 \cite{mclean1971rabies}.
	\par  In contrast, in the case of {\it IPP} arrangement of initial infection, the disease also spreads, in the beginning, with migration into neighbouring infection-free patches of the lattice as shown in the panels (B1)-(B4) in Fig.~\ref{snap_grid}, however, a recovery process starts  with higher migration and finally the lattice becomes infection-free for larger migration (panel B5, Fig.~\ref{snap_grid}, $\epsilon=0.65$).  
	\par For the initial {\it RDIP} distribution of infected sites, the spreading of infection and its eradication  follow a trend similar to the {\it IPP} case. The spreading process occurs in a comparably smaller range of migration rate as seen in the lower panels (C1)-(C4) in Fig.~\ref{snap_grid}, however, the lattice also becomes infection-free for a smaller migration rate (panel C5, Fig.~\ref{snap_grid}) compared to the {\it IPP} case. The infection spreads very fast to cover almost $80\%$ of the patches at a  migration rate $\epsilon=0.04$ and it is eradicated at a relatively faster rate to become infection-free at a lower $\epsilon=0.316$.
	
	\subsection{Combined role of migration and force of infection}
\begin{figure*}[!h]
		\centering
		\includegraphic[height=4cm,width=4.4cm]{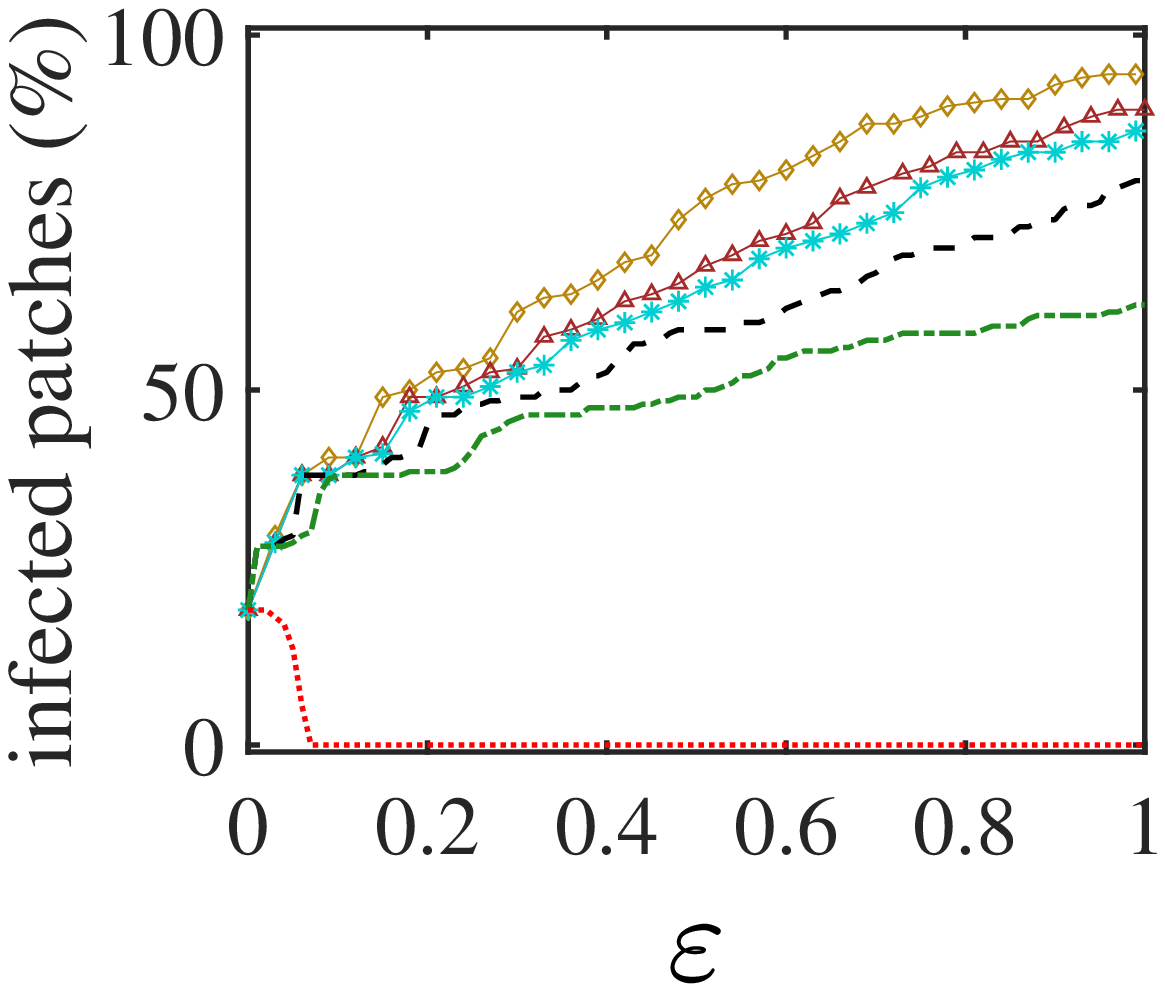}{(a) \quad~~{\it ICP}}{35pt}{-0.0}
		\hspace{3pt}\includegraphic[height=4cm,width=4.2cm]{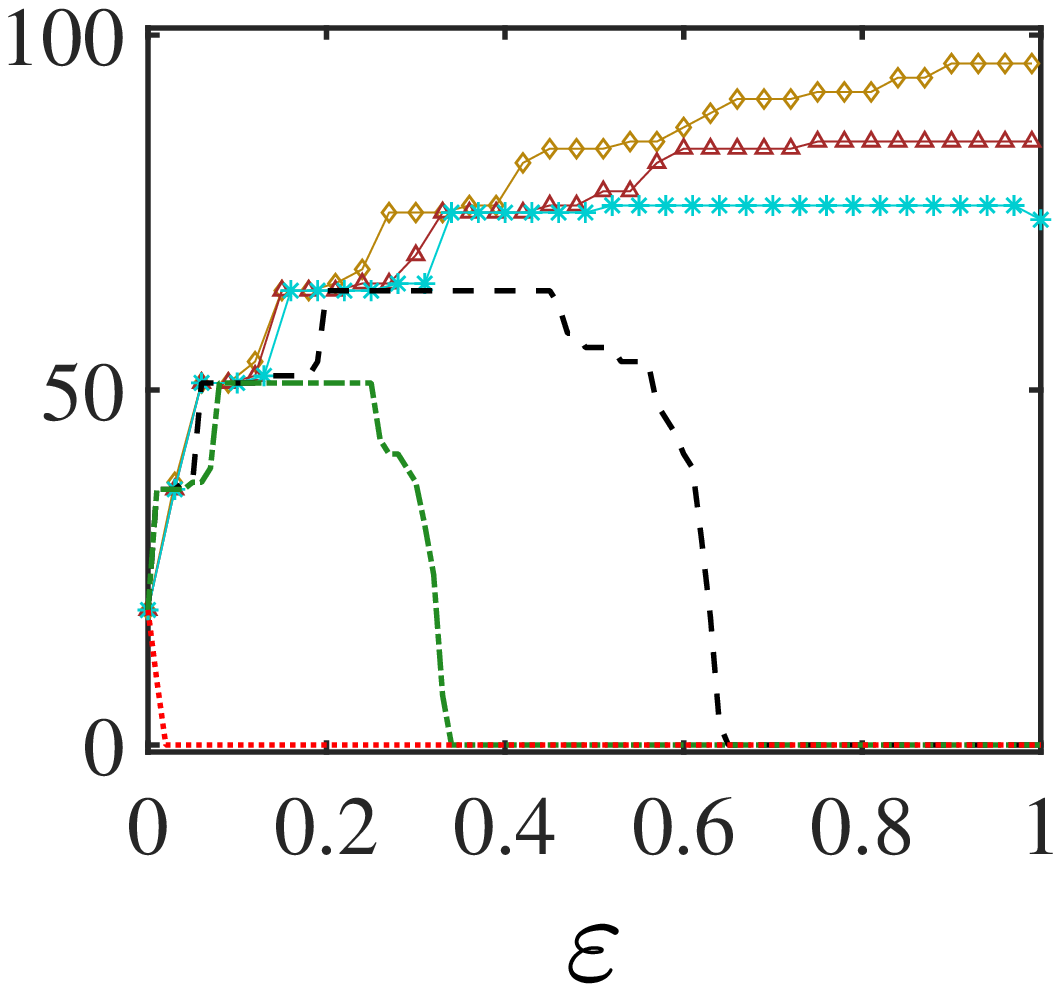}{(b)\quad \quad {\it IPP}}{25pt}{-0.0}
		\hspace{5pt}\includegraphic[height=4cm,width=4.2cm]{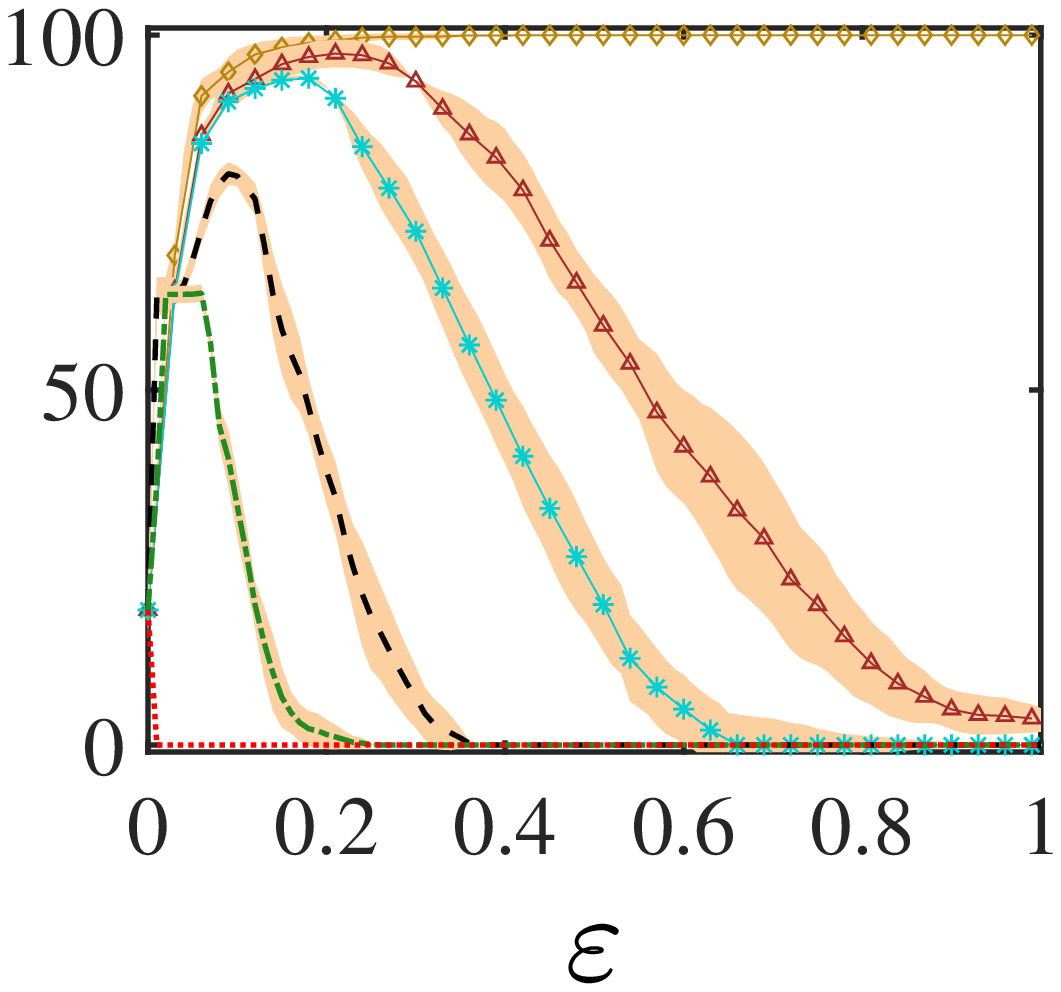}{(c)\quad \quad {\it RDIP}}{25pt}{-0.0} \\
		\includegraphic[height=4cm,width=4.5cm]{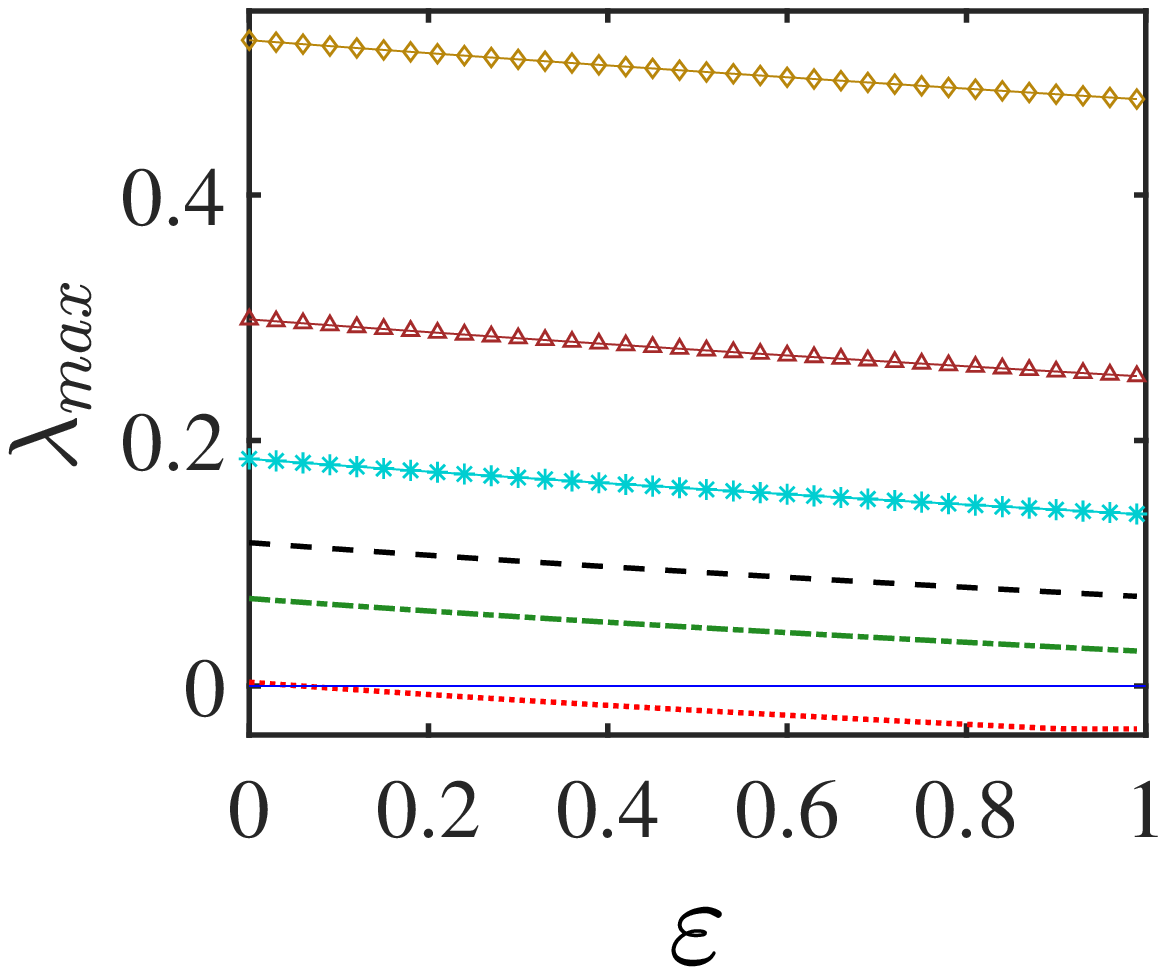}{(d)}{110pt}{1.2}
		\includegraphic[height=4cm,width=4.4cm]{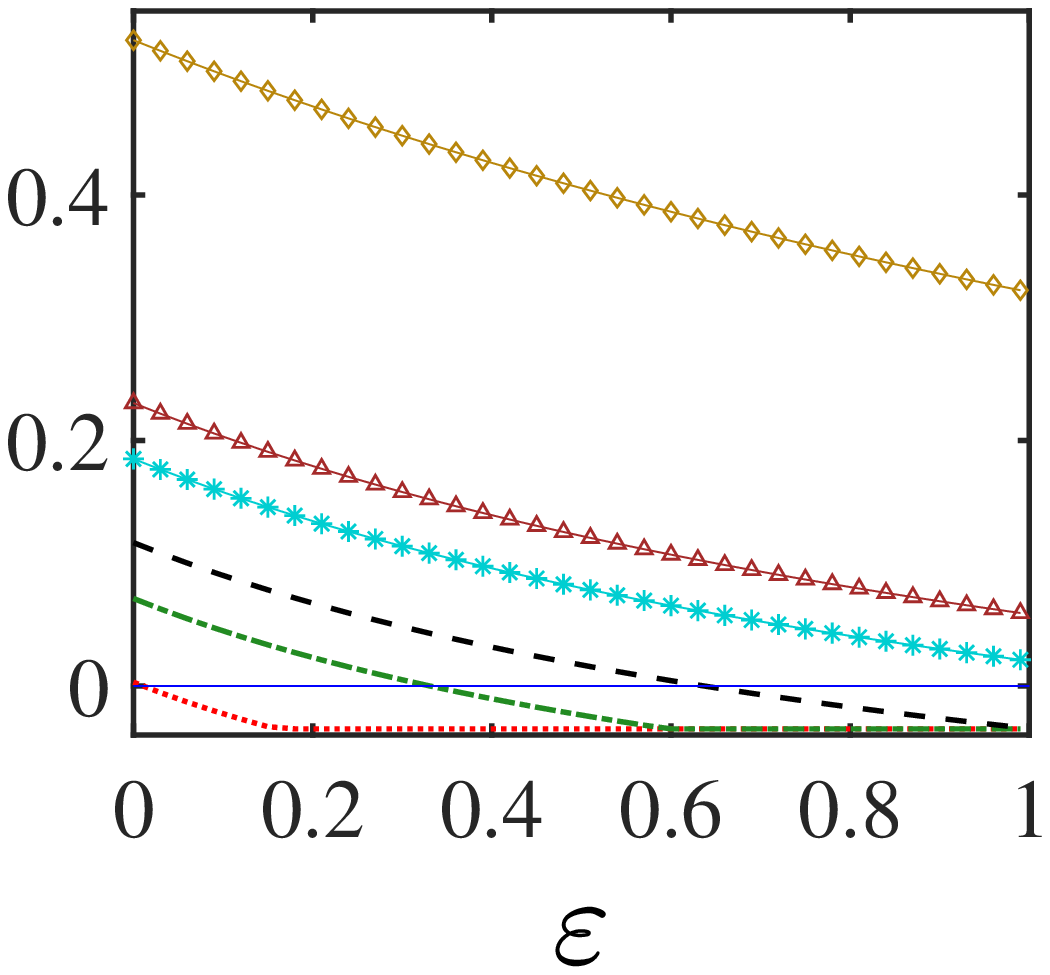}{(e)}{105pt}{1.2}
		\hspace{-2pt}\includegraphic[height=4cm,width=4.4cm]{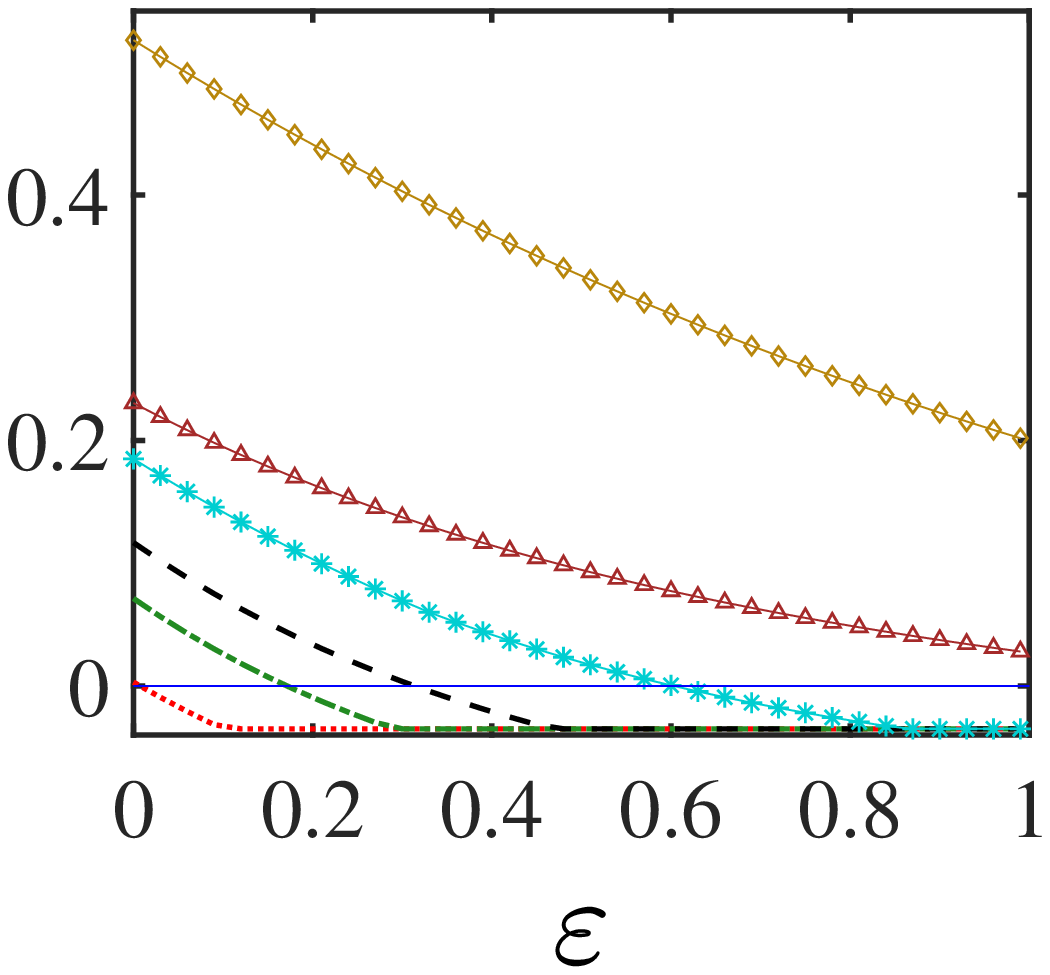}{(f)}{105pt}{1.2}\\
		\hspace{10pt}\includegraphic[height=4cm,width=4cm]{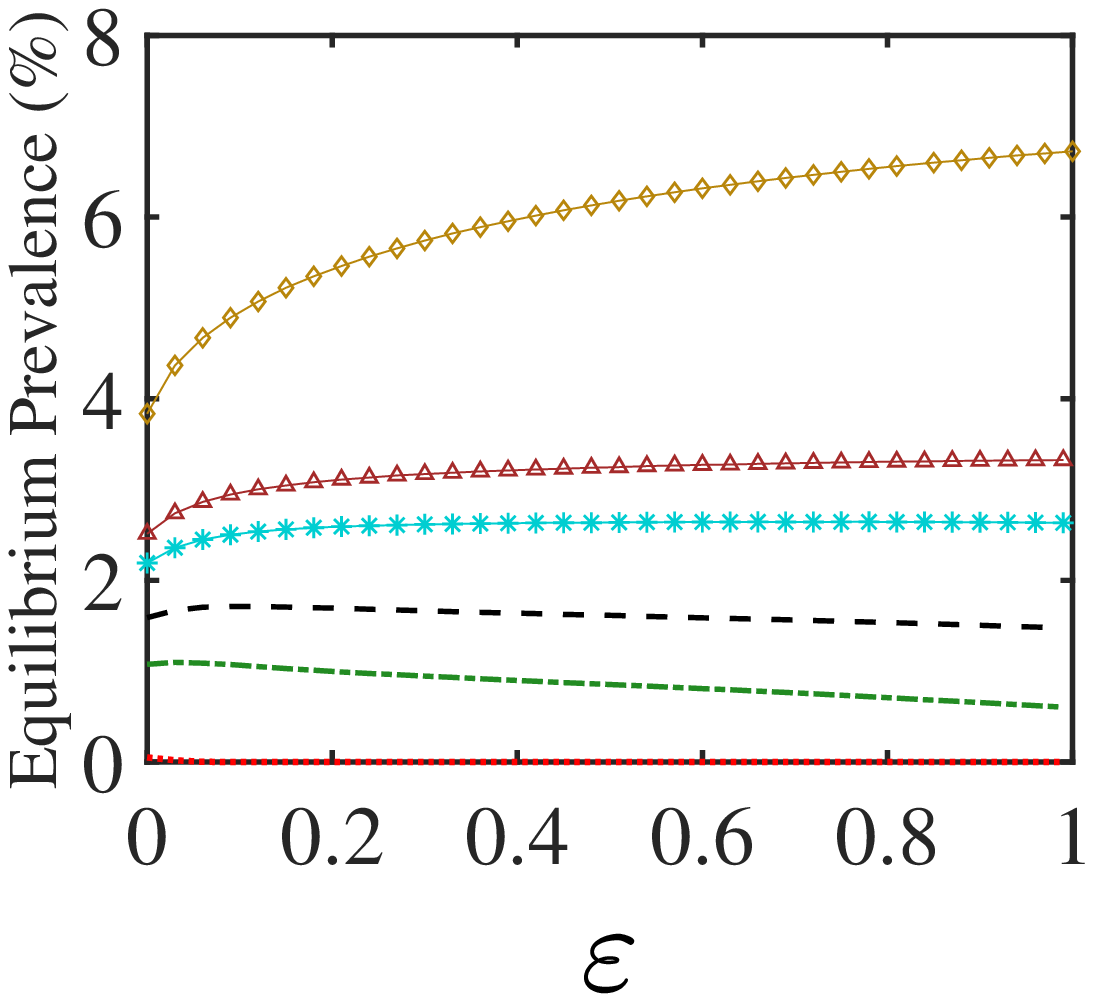}{(g)}{95pt}{1.2}
		\hspace{13pt}\includegraphic[height=4cm,width=4cm]{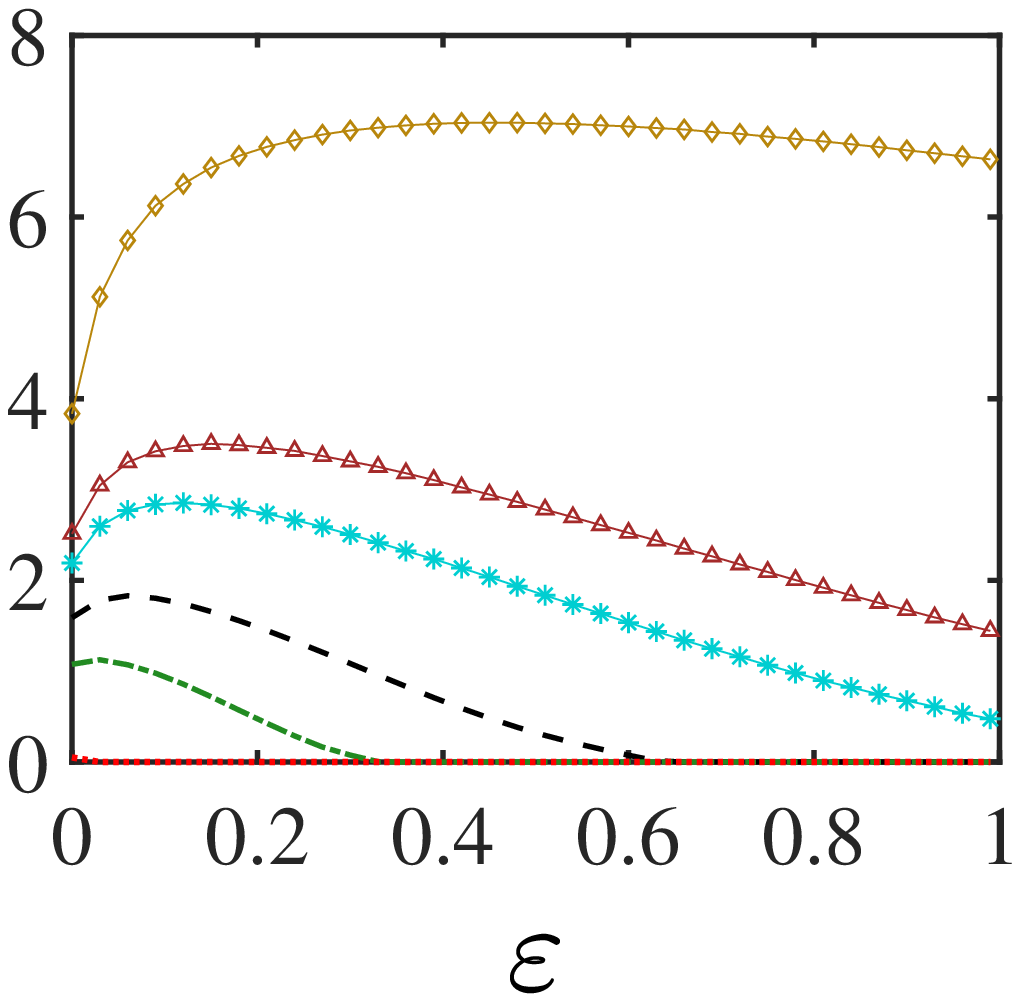}{(h)}{95pt}{1.2}
		\hspace{10pt}\includegraphic[height=4cm,width=4cm]{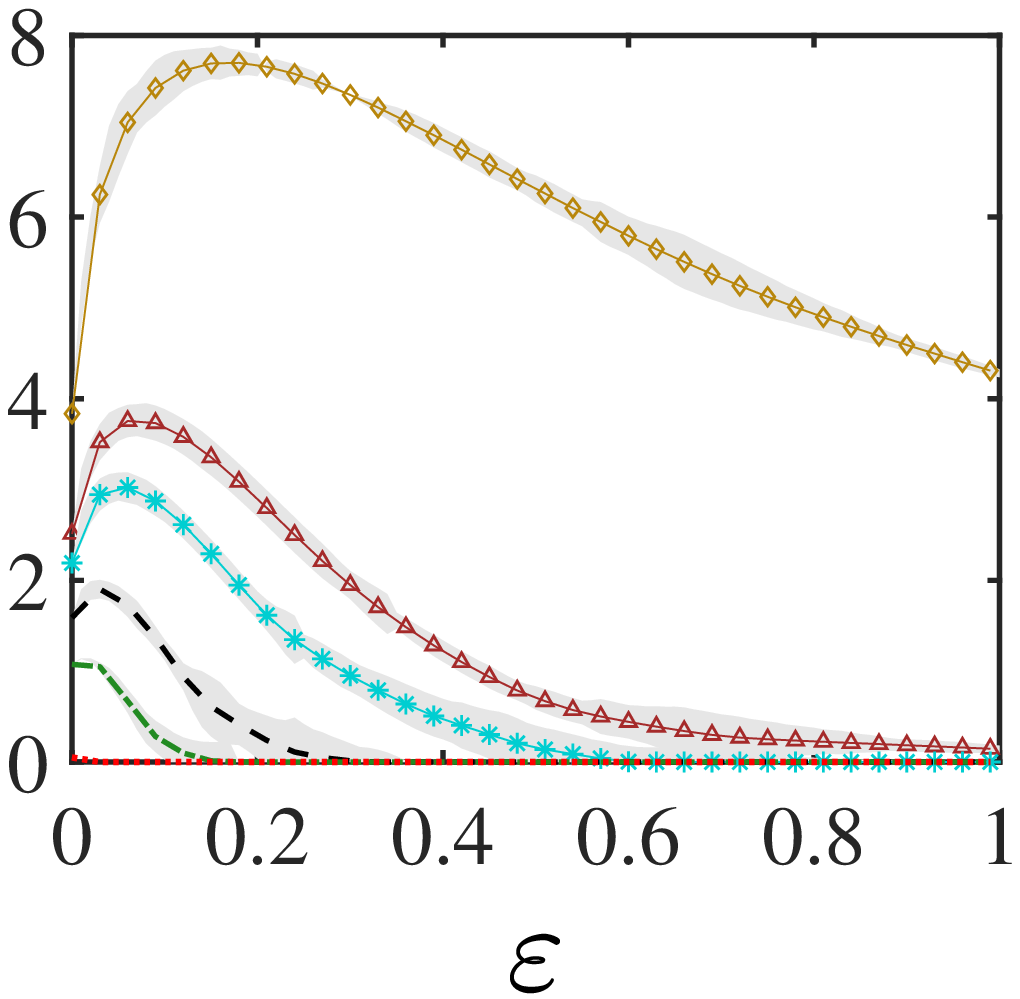}{(i)}{95pt}{1.2}
		\includegraphics[height=0.25in,width=5in]{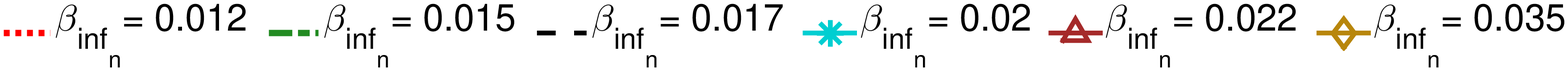}
		\caption{Upper row: Progress of infection and recovery process in the lattice for varying migration rate, $\epsilon$. Each of the (a) {\it ICP}, (b) {\it IPP} and (c) {\it RDIP} distribution starts with $19\%$ infected patches and five different force of infections $\beta_{inf_n}$ = (0.012, 0.015, 0.017, 0.02, 0.022, 0.035). The critical migration rates for which the lattice is globally recovered with a given force of infection are as follows: (a) {\it ICP} case, $\epsilon_c=0.07$ for $\beta_{inf_n}=0.012$; (b) {\it IPP} case, $\epsilon_c$ = (0.02, 0.34, 0.65) for $\beta_{inf_n}$ = (0.012, 0.015, 0.017), respectively; (c) in {\it RDIP} case, $\epsilon_c$ = (0.01, 0.2, 0.35, 0.66) for $\beta_{inf_n}$ = (0.012, 0.015, 0.017, 0.02), respectively. The shaded regions indicate standard deviations. It shows that increasing migration can make the lattice disease-free for various $\beta_{inf_n}(>\beta_c$) in case of {\it RDIP}, but remains unsuccessful in case of {\it ICP} and {\it IPP}. Middle row: (d-f) Largest eigenvalues ($\lambda_{max}$) are plotted using Eqs.~\eqref{jac2}-\eqref{eq5} to confirm stability of the HSS when the critical migration rate  crosses the zero line (solid blue line). Such critical migration rates are estimated for all three cases as: ICP case, $\epsilon_c=0.07$ for $\beta_{inf_n}=0.012$;  IPP arrangement, $\epsilon_c$ = (0.02, 0.34, 0.65) for $\beta_{inf_n}$ = (0.012, 0.015, 0.017), respectively; RDIP case,  $\epsilon_c$ = (0.01, 0.2, 0.35, 0.66) for $\beta_{inf_n}$ = (0.012, 0.015, 0.017, 0.02), respectively.	
			Lower row: Percentage of equilibrium prevalence of disease in the lattice for varying migration rate corresponding to each $\beta_{inf_n}$ for (g) {\it ICP}, (h) {\it IPP} and (i) {\it RDIP} distributions. Parameters are as given in  Fig.~\ref{fig2}, particularly, $\beta_{o_m}=0.002$ is kept always fixed. }\label{all} 
	\end{figure*}

	A broader scenario of infection spreading and recovery processes is presented here  (Fig.~\ref{all}) for all three initial arrangements of infected sites under the joint effect of migration and  a range of forces of infection, $\beta_{inf_n}$ = (0.012, 0.015, 0.017, 0.02, 0.022, 0.035), when  $19\%$ patches of the lattice are initially infected  ($81\%$ infection-free patches). In the case of {\it ICP} arrangement, the number of infected patches increases almost monotonically with  migration $\epsilon$ (Fig.~\ref{all}(a)). The lattice becomes infection-free only for a very low  force of infection $\beta_{inf_n}=0.012$ or less, otherwise, the lattice never recovers from the infection.
	On the other hand, for initial {\it IPP} arrangement (Fig.~\ref{all}(b)), the number of infected patches reaches a maximum level and then starts declining for a range of forces of infection, $\beta_{inf_n}$ = (0.012, 0.015, 0.017) and the lattice eventually recovers from the disease  at their respective critical migration rates ($\epsilon_c$ = 0.02, 0.34, 0.65). The lattice never becomes infection-free in the range of larger  $\beta_{inf_n} (=0.02, 0.022, 0.035)$. In the case of {\it RDIP} distribution (Fig.~\ref{all}(c)), the spreading and the recovery processes follow a similar trend as observed in the  {\it IPP} case, however, the disease eradication process is significantly enhanced. The  lattice becomes infection-free at lower critical values of migration, $\epsilon_c$ = (0.01, 0.2, 0.35, 0.66) for the respective forces of infection, $\beta_{inf_n}$ = (0.012, 0.015, 0.017, 0.02) considered for the other two cases. For a larger $\beta_{inf_n}=0.022$, the global recovery does not occur, but a trend of significant recovery is noticed with a largely decreasing number of infected patches. However, if infection pressure is too high ($\beta_{inf_n} \geq 0.035$), even a higher rate of migration fails to start a recovery process and a global prevalence (infection spreads into all the patches of the lattice) occurs at a significantly lower value of $\epsilon=0.44$. The shaded region in each plot indicates the standard deviation estimated from average values of 50 numerical realizations by varying only the random locations of the infected sites, while other parameters and conditions are kept fixed. Hence results of the {\it RDIP} initial distribution are robust, in the sense, that the qualitative nature remains unaffected by the changes in random positions of the infectious patches. We observe  a dual nature of spreading and healing processes for {\it RDIP} initial distribution  of infection under migration in a global recovery of the lattice.
	The infection initiated from random sites can enhance resilience under low and moderate forces of infection, however, it is less resistive to the considerably high virulence of attack, hence cannot prevent a global outbreak. Other two spatial arrangements are ({\it ICP, IPP}) although vulnerable against the epidemic that promotes spreading, the percentage of infected patches is less compared to the {\it RDIP} case. As mentioned in Sec.~III, results are qualitatively independent of the threshold level of infected hosts such as $y_h=0.001$, as illustrated in Figs.~\ref{all2}(a)-\ref{all2}(c) (see Appendix A).
	
	\par We interpret our results from a purely dynamical system point of view, using the pictures of the time evolution of infected and infection-free hosts in all the sites for increasing migration for all three initial distribution patterns of infected sites in the lattice. For the initial {\it ICP} arrangement, we can explain the dynamics of the lattice as a transition  to inhomogeneous steady-states (IHSS) \cite{hens2013oscillation, koseska2013transition, hens2014diverse,nandan2014transition} for increasing migration in the lattice. In the IHSS, the lattice is stabilized to different steady-state values (infected or infection-free) as shown in the time evolutions presented in the Appendix B (Figs.~\ref{time_ABC45}(a)-(b) and Figs.~\ref{time_ABC45}(c)-(d) for $\epsilon=0.5$ and $\epsilon= 1$, respectively, for both the hosts $x_i$ and $y_i$). A large number of sites
		coexists with both hosts ($x_i> 0, y_i> 0$) and some others exist with susceptible hosts only ($x_i> 0, y_i= 0$). The lattice with initial {\it ICP} distribution of infection never realizes a completely infection-free state ($x_i> 0, y_i= 0$) for larger migration. 
		For the IPP case, the time evolution for increasing migration rate show a gradual convergence to a homogeneous steady state (HSS) (see, Appendix B, Figs.~\ref{time_ABC45}(e)-(f) and Figs.~\ref{time_ABC45}(g)-(h)). 
		In this case, $x_i$ and $y_i$ converge to a unique steady-state, which defines the desired infection-free equilibrium ($x_i=x_s, y_i=0)$ for $\epsilon=0.65$.  
		For the {\it RDIP} initial arrangement, the lattice clearly passes through an intermediate IHSS for $\epsilon=0.2$ to a final infection-free HSS state ($x_i=x_s, y_i=0$) at $\epsilon=0.316$ (see, Appendix B, Figs.~\ref{time_ABC45}(i)-\ref{time_ABC45}(l)). 
		\par  Noteworthy that we do not observe any oscillatory states for our selected range of parameters in the whole investigation as seen in Fig.\ref{time_ABC45}. We always dealt with either the IHSS or HSS. The local stability of such a state as our most desired infection-free equilibrium  $E_s=(x_s,0)$ is checked for all the three cases. We derive the {\it Jacobian} matrix by linearizing the whole system at the infection-free equilibrium (see Appendix C) and then plot the largest eigenvalues as a function of migration rate $\varepsilon$ using Eqs.~\eqref{jac2}-\eqref{eq5} for all the three cases, {\it ICP}, {\it IPP} and {\it RDIP} in Figs.~\ref{all}(d)-\ref{all}(f)). The largest eigenvalue crosses the zero value (blue horizontal line) at a critical value $\epsilon=\epsilon_c$, where the lattice emerges with a stable infection-free state.  By comparing  Figs.~\ref{all}(a)-\ref{all}(c) and Figs.~\ref{all}(d)-\ref{all}(f), it is confirmed that in all the three cases, $\epsilon_c$ values   exactly match with their numerically computed $\epsilon_c$ values for respective $\beta_{inf_n}$ values. Especially to mention that for the {\it ICP} case, the largest eigenvalue crosses the zero line (blue line) only for a low $\beta_{inf_n}=0.012$, thereby confirming the worst nature of recovery in this case.
	\par 
	We also determine the disease prevalence, an important metric in epidemiology that defines the fraction of infected individuals. It measures the severity of the disease and quantifies the chance of disease transmission. The larger the value of prevalence, the greater is the chance of disease transmission 
	\cite{holt2007predation}. For a single patch, prevalence is determined by the ratio of infected hosts to the total population. The fraction of infected population  population in the lattice at an equilibrium state (Equilibrium Prevalence $\%$) is measured by,
	\begin{align}
	\text{Equilibrium Prevalence} (\%) = EPP =
	\frac{1}{N\times N}\sum_{i=1}^{N\times N} \frac{\tilde{y}_i}{\tilde{x}_i+ \tilde{y}_i}\times 100,		
	\end{align} 
	where $\tilde{x}_i$ and $\tilde{y}_i$ are, respectively, the steady-state values of susceptible and infected populations of the $i^{th}$ patch. If all the patches in the lattice are infection-free then $EPP=0$, otherwise $EPP\neq 0$.
	{\it EPP} of the entire lattice for a given force of infection steadily increases with increasing migration in the case of {\it ICP} arrangement for moderate to higher values of $\beta_{inf_n} $ (Fig.~\ref{all}(g)), whereas it decreases  for {\it IPP} and {\it RDIP} distributions (Figs.~\ref{all}(h) and \ref{all}(i)). It presents a comparative scenario by the fraction of infected population. More than $7.5\%$ of the total hosts at equilibrium appears infected across the lattice at a lower migration rate $\epsilon=0.175$ for the {\it RDIP} arrangement (Fig.~\ref{all}(i)), however, it is much lower in {\it ICP} and {\it IPP} cases (Figs.~\ref{all}(g) and \ref{all}(h)) at the same migration rate. Though the infection declines at a faster rate with increasing migration (for all the $\beta_{inf_n}$ values cited here) in {\it RDIP} arrangement, effectively eradicating the infection for low to moderate values of migration.
	
	\subsection{Effect of initial number of infected patches}
\begin{figure}[H]
		\centering
		\includegraphic[height=5cm,width=5.25cm]{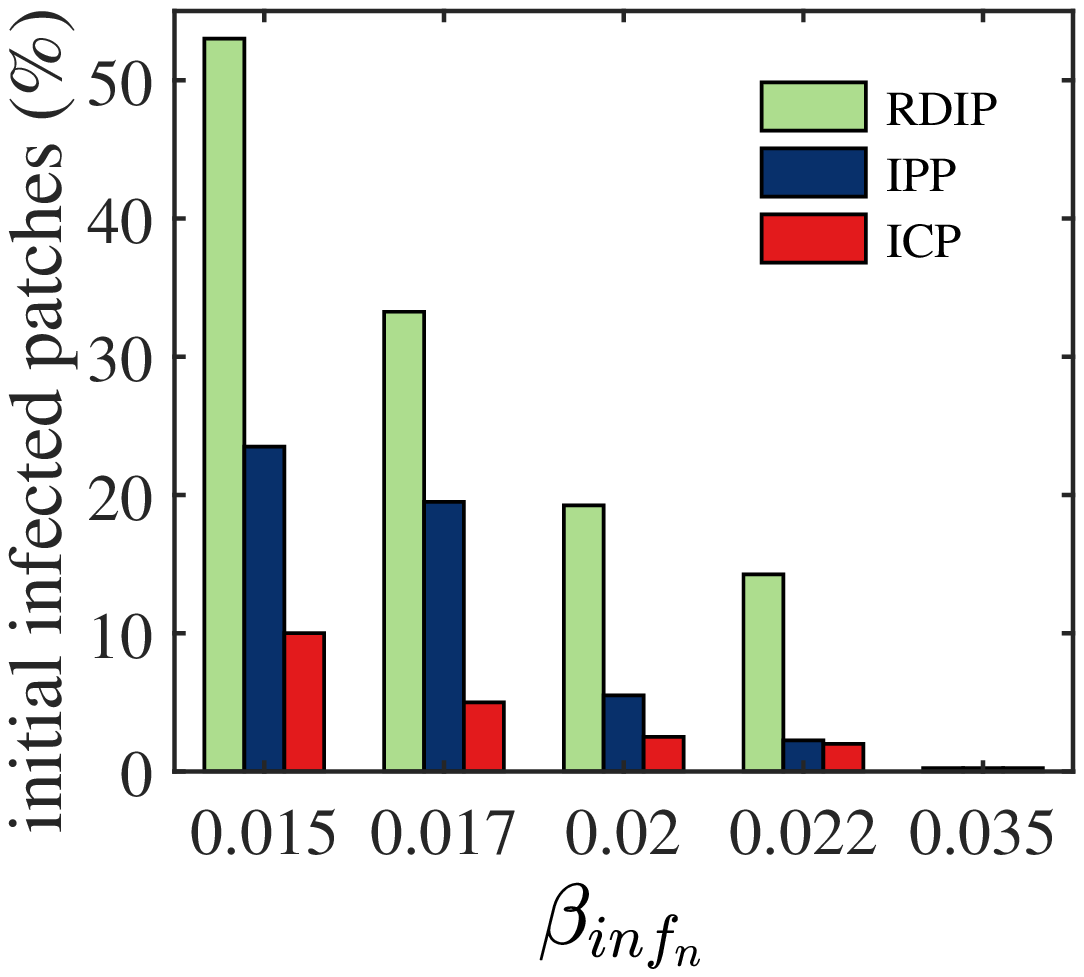}{(a)}{0pt}{-0.3}
		\hspace{20pt}
		\includegraphic[height=5cm,width=5.45cm]{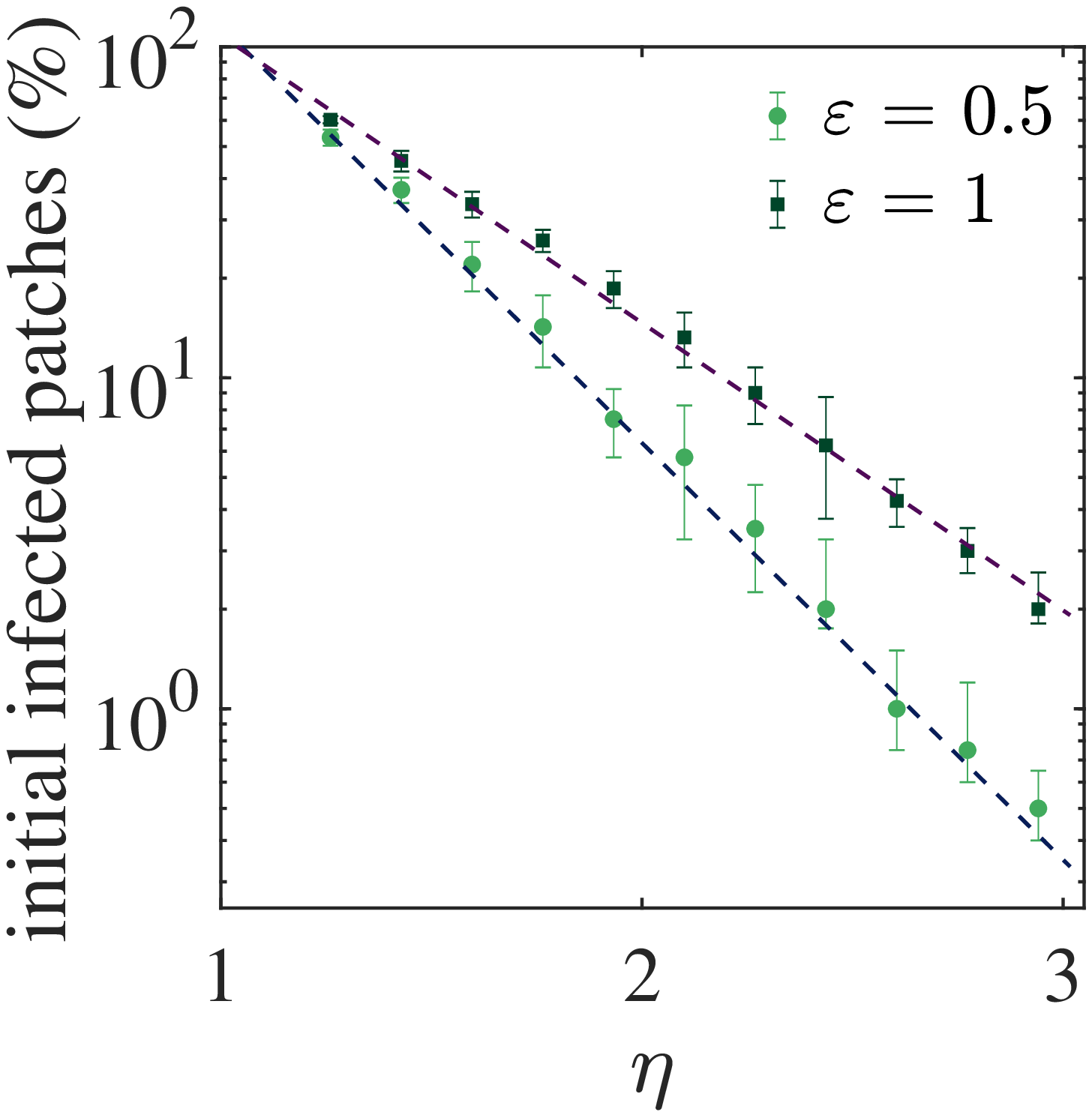}{(b) \quad \quad \quad \quad \quad RDIP }{0pt}{-0.3}
		\caption{(a) Recovery limit of the lattice under {\it ICP, IPP} and {\it RDIP} initial distributions of infected sites. The percentage of initial infected patches from which a lattice can recover (a global recovery) for different forces of infections ($\beta_{inf_n}=0.015,0.017,0.02,0.022,0.035$). Parameters are as in Fig.~\ref{all} with $\epsilon=1$. (b) Initial infected patches (\%) and the relative infectivity, $\eta=\beta_{inf_n}/\beta_c$ follows an exponential law in the disease eradication process for  the {\it RDIP} case. Dashed lines are the best fitting for exponential decay of initial infection (\%) for two selected migration rates $\epsilon=0.5$ (green circles) and $\epsilon=1$ (black squares). Error bars indicate standard deviations from an average of 50 individual simulations by varying the initial positions of infected patches only. }\label{prob}
	\end{figure}

	Results of disease spreading and recovery are so far presented with $19\%$ initial infected patches. Now we check how does an arbitrarily chosen larger fraction of initially infected patches affect the recovery process? Figure~\ref{prob}(a) presents the overall scenario of disease spreading and the recovery process, for all the three initial arrangements, where we vary the initial  fraction of infected sites.
	It reconfirms that the lattice has a better ability, in the case of {\it RDIP} initial arrangement (green bars), to heal the entire lattice even from a situation of a larger fraction of initially infected patches. Results are repeated for five different forces of infection $\beta_{inf_n}$ (0.015, 0.017, 0.02, 0.022, 0.035) at a fixed $\epsilon=1$ and compared with the  other two cases, $ICP$ and $IPP$ arrangements (blue and red bars). For example, for a $\beta_{inf_n}=0.015$ and  {\it RDIP} arrangement, the lattice is able to recover from initial infection in more than $50\%$ of patches, while for {\it ICP} and {\it IPP} arrangements, it cannot recover from such a large fraction of initially infected patches. The lattice, with {\it IPP} initial arrangement, can recover at best from initial infection of about one-fourth of its patches (blue bar in Fig.~\ref{prob}(a)) for a moderate value of $\beta_{inf_n}$=0.015.  The  recovery process is exceptionally poor for initiation of the disease from centrally located patches ({\it ICP} arrangement); at best it can recover from an initial infection of about 10$\%$ of patches (red bar) for $\beta_{inf_n}=0.015$. Compare with Fig.~\ref{all}(a), for {\it ICP} arrangement, the lattice can recover with 19\% initial infected patches for $\beta_{inf_n}=0.012$ only. 
	If the transmission rate $\beta_{inf_n}$ is further increased, the recovery process declines to a lesser number of initially infected sites (\%) for all the three cases, however, the {\it RDIP} case always outperforms the other two initial  spatial arrangements. 
	Our results indicate that infection management would be harder for both the {\it ICP} and {\it IPP} cases for infectious diseases with high transmissibility. Since in the case of {\it RDIP} initial distribution, the lattice can handle a larger fraction  of initial infected patches, we explore the situation further with a wide variation of the force of infection.
	For a better presentation of the scenario, we redefined the force of infection as relative  infectivity with a  normalization of $\beta_{inf_n}$ by the critical value $\beta_c$ of $\beta$ above which a disease can invade a single patch. Thus the relative infectivity, $\eta=\frac{\beta_{inf_n}}{\beta_c}$, is used in search of the underlying  recovery process for this {\it RDIP} distribution. Noticeably, for increasing $\eta$, the lattice follows the exponential decay of initial fraction  of infected patches from which the lattice can start a self-organized process of recovery, as shown for two different migration rates $\epsilon=0.5$ (green circles) and $1.0$ (black squares) in Fig.~\ref{prob}(b). More accurately, the recovery process follows an exponential law defined by $\tau e^{-\eta\xi}$, where $\tau$ and $\xi$ are positive parameters that measure the slope of the fitted lines (dashed lines) in a semi-log scale. The slope of the exponential line varies with the rate of migration. The lattice can be globally recovered if initially infected sites are below the line of the corresponding migration rate. Chances of recovery decrease with increasing order of relative infectivity, $\eta$. Similar investigations for two other initial spatial arrangements of infection also show similar trends, however, their percentages of initial infections that start a self-organizing recovery process, goes so low with relative infectivity that we do not want to make any conclusive statement at this moment. 
	
	\subsection{ Recovery process: Role of random rewiring}
\begin{figure*}
		\centering	
		\includegraphic[height=3.5cm,width=3.6cm]{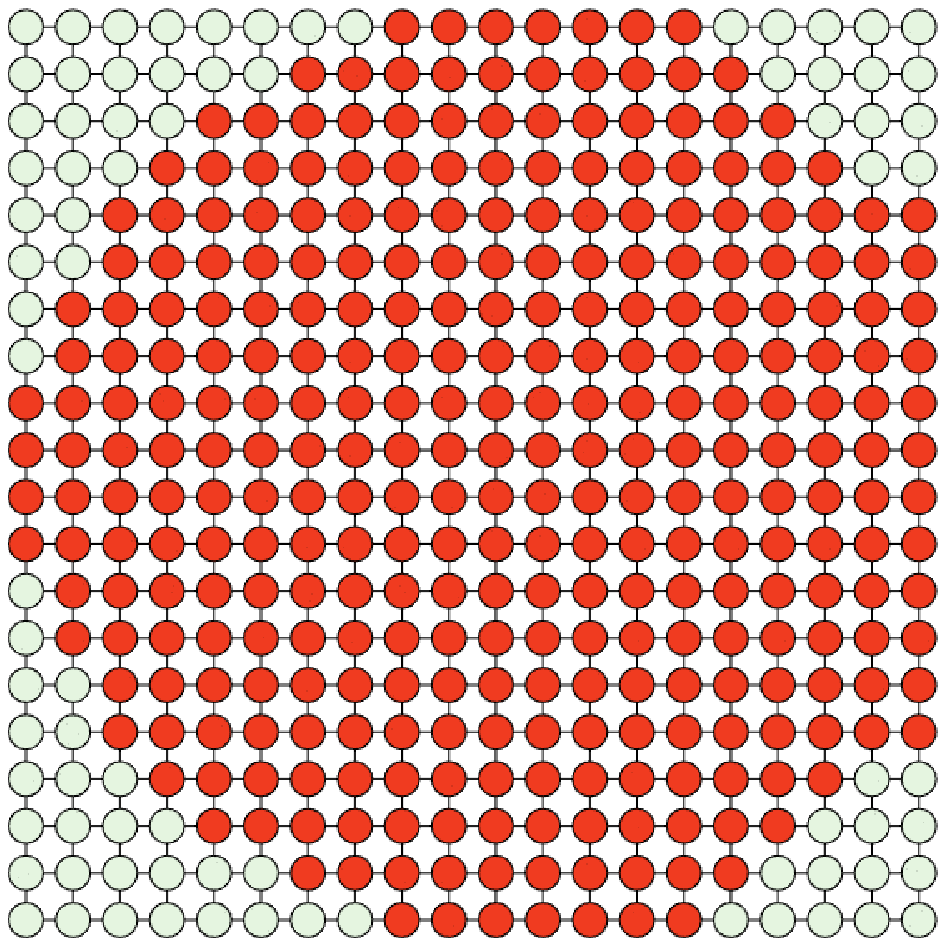}{(a)  $p=0$}{0pt}{0}
		\includegraphic[height=3.5cm,width=3.6cm]{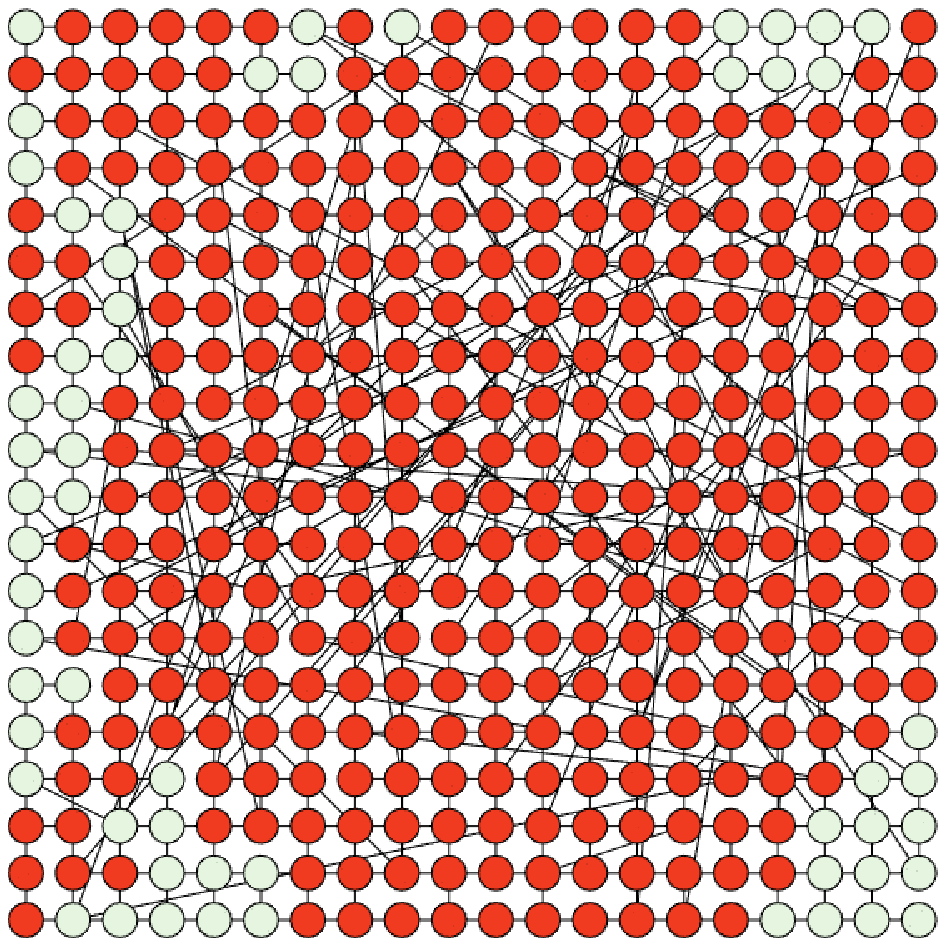}{(b)  $p=0.1$}{0pt}{0}
		\includegraphic[height=3.5cm,width=3.6cm]{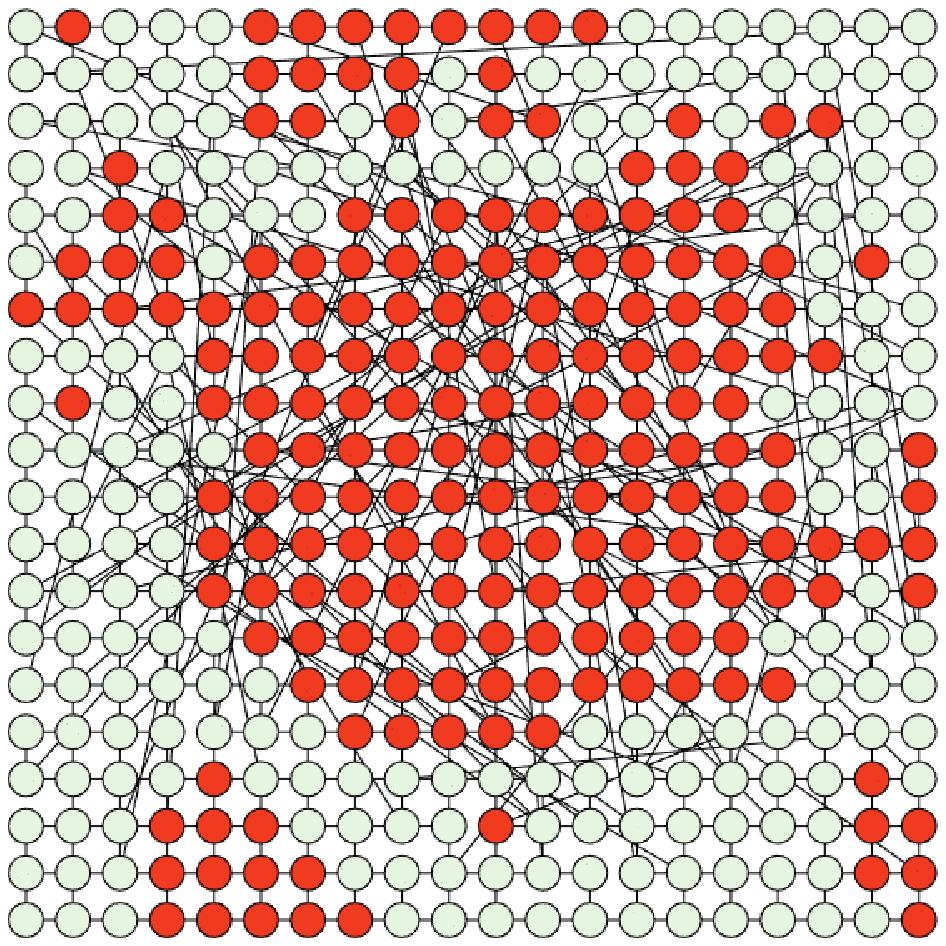}{(c)  $p=0.15$}{0pt}{0}
		\includegraphic[height=3.5cm,width=3.6cm]{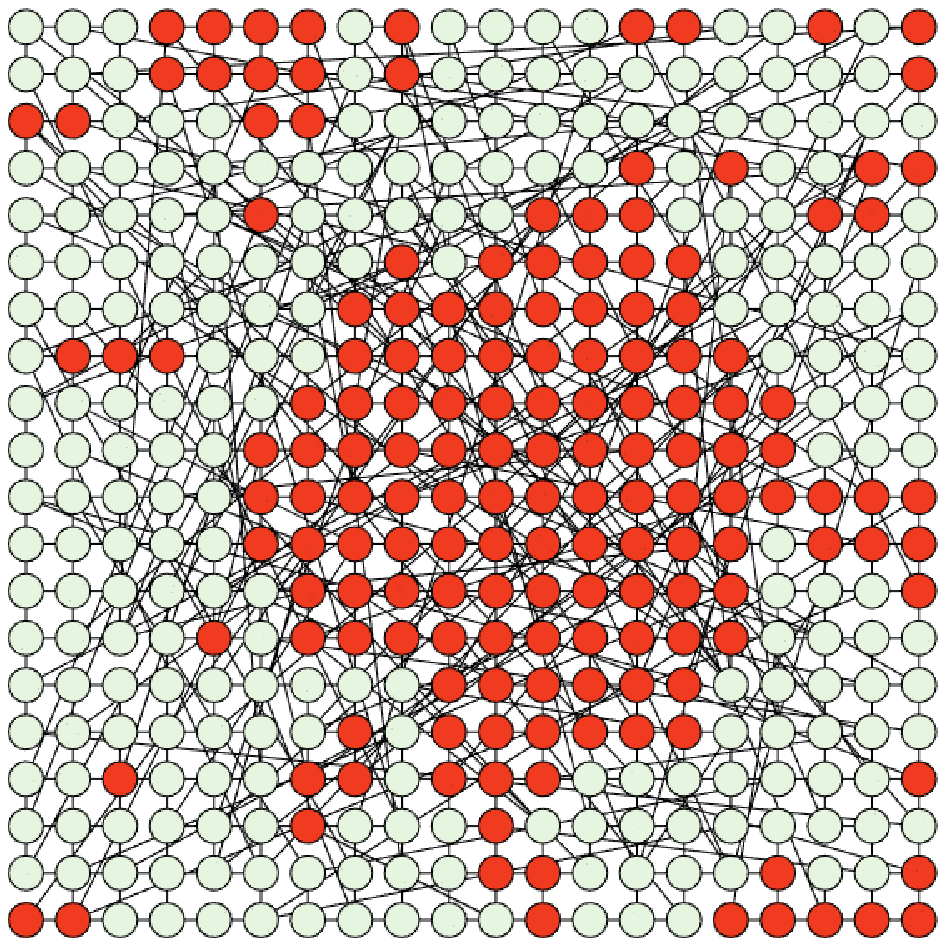}{(d) $p=0.2$}{0pt}{0}\\
		\includegraphic[height=4.25cm,width=4.5cm]{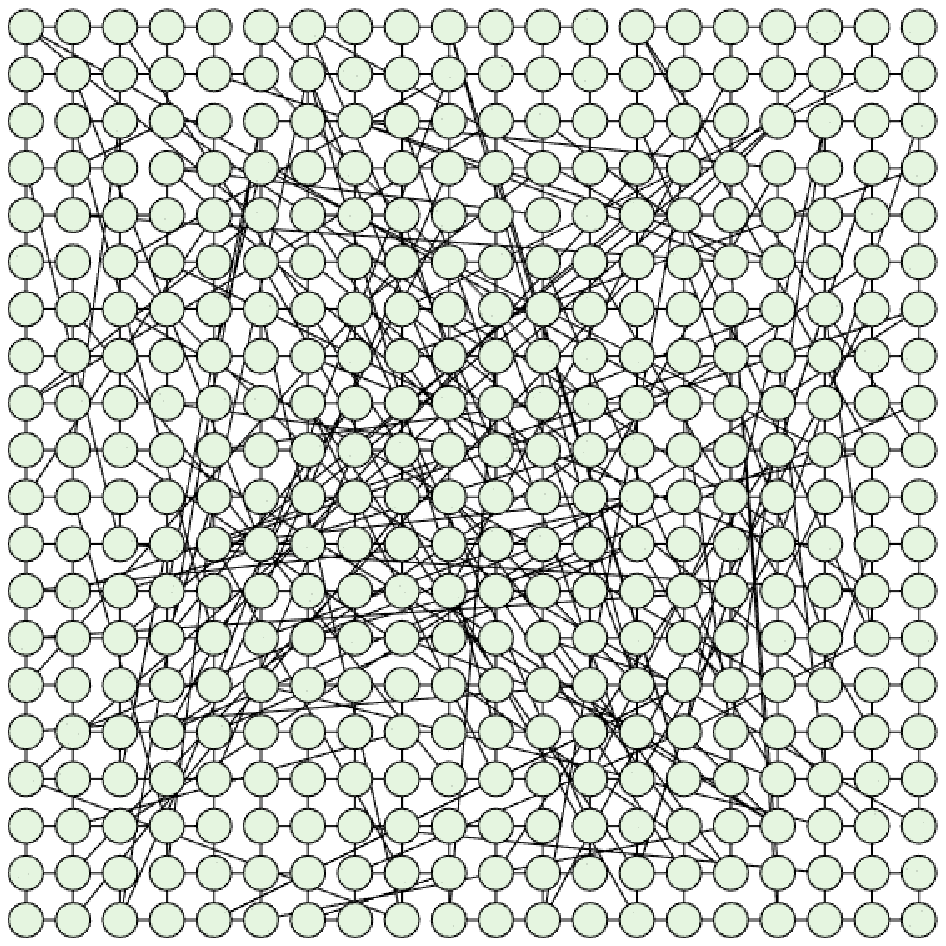}{(e) $p=0.22$}{0pt}{0}
		\includegraphic[height=5.25cm,width=5.25cm]{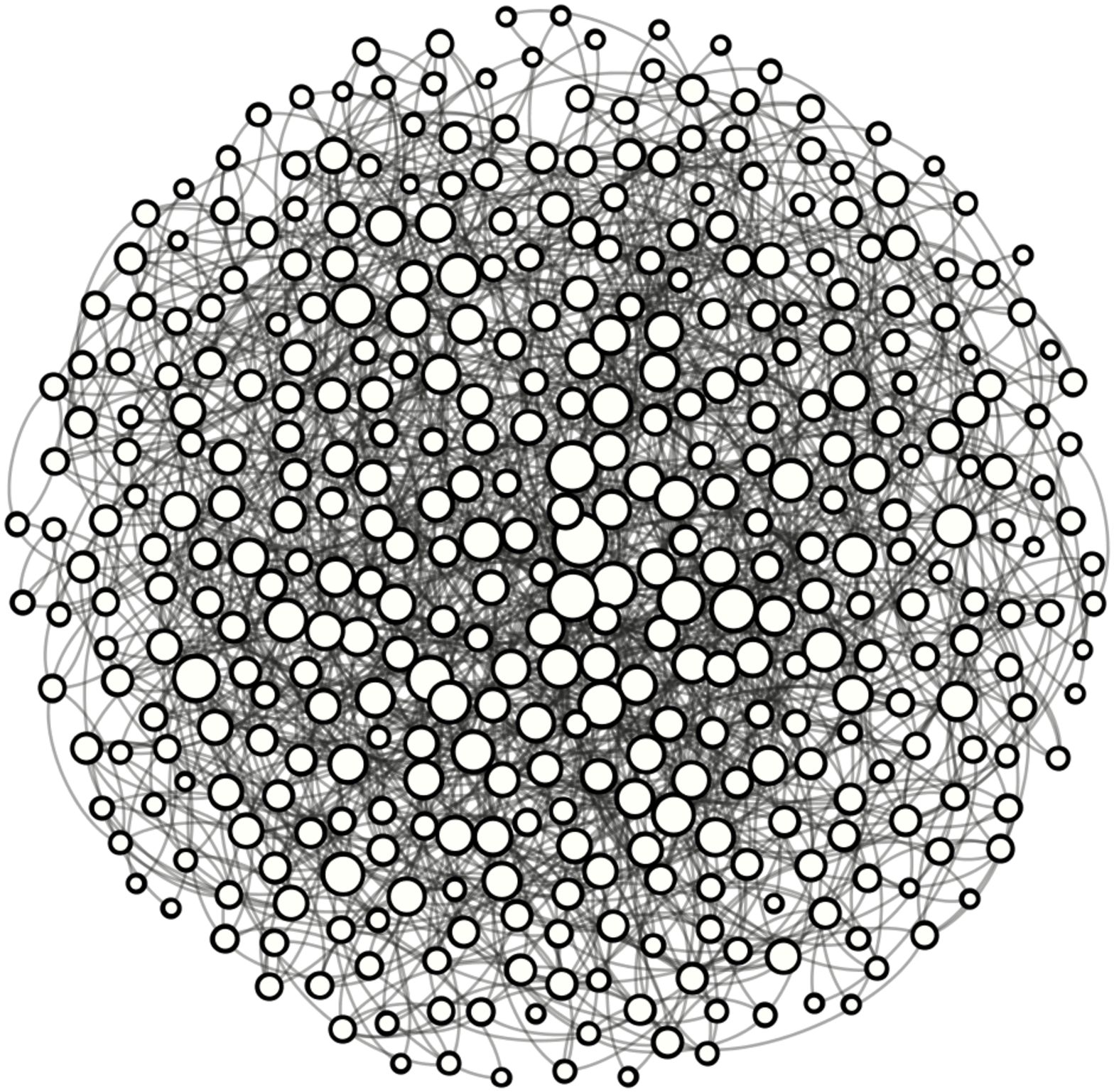}{(f) $p=0.22$}{0pt}{1.4} 
		\caption{Snapshot of infected patches in the lattice with increasing rewiring probability ($p$) in case of {\it ICP} initial arrangement of infection. (a) Initial infection in $75\%$  of the patches at  a fixed migration rate $\epsilon=1$ and $\beta_{inf_n}=0.017$ where $p=0$ (similar to Fig.~\ref{snap_grid}(A5)), (b) infection spreads to more  infection-free patches when few long range connections   ($p=0.1$) are  added. Recovery starts with a gradual decrease in the number of infected patches for increasing rewiring probability (c) $p=0.15$, (d) $p=0.2$. (e) A new network emerges from the lattice that becomes infection-free at $p=0.22$. Long-range connections are indicated by thin black lines in (b)-(e).  
				(f) Final emergent network ($p=0.22$) that presents a better visualization of degree distributions of each node or patch, where the size of the circles indicates the degree of the nodes. Other parameters are the same as given in Fig.~\ref{snap_grid}(A5). }\label{rewir_snap}
	\end{figure*}

	We focus here on the case of {\it ICP} initial arrangement that is the worst performer in the recovery process of the lattice. The randomization of the initial spread of infected sites, as adopted for the {\it RDIP} case, significantly improves the recovery process of the lattice, which encourages us to apply a kind of randomness in the {\it ICP} arrangement by rewiring of links of the lattice with some probability, $p$, keeping the total number of links unchanged. The rewiring strategy \cite{sw} allowed us to construct new  connectivity graph from regular ($p=0$) to random ($p=1$) by introducing long- and short-range interactions among the patches that showed a history of improved performance with an emergent network. The movement of susceptible and infected hosts is no longer restricted to nearest neighbours only, hosts are rather allowed by this rewiring to migrate to long-distance sites of the lattice. We initiated the study with $85\%$ infected patches in {\it ICP} initial arrangement at a migration rate $\epsilon=1$, $\beta_{inf_c}=0.017$ and $p=0$ as shown in Fig.~\ref{rewir_snap}(a) (which is the original lattice structure Fig.~\ref{snap_grid}(A5) with no rewiring). A random rewiring adds new direct connections or links (thin black lines) between the long-distance patches beyond immediate neighbors and also withdraws some of the nearest neighbor links. As a consequence of our proposed rewiring, infection spreads into non-infected patches, and at first deteriorates the epidemic situation in the emerging networks (Fig.~\ref{rewir_snap}(b)) for varying $p$. However, a gradual recovery of infected patches in the lattice is then noticed as the rewiring probability is further increased as shown in Figs.~\ref{rewir_snap}(c) and \ref{rewir_snap}(d). A global recovery from infection is observed at a rewiring probability $p\geqslant0.22$ (Fig.~\ref{rewir_snap}(e)). The emergent  network structure (with rewired black links) in Fig.~\ref{rewir_snap}(e) for $p=0.22$ is redrawn in Fig.~\ref{rewir_snap}(f) for a better visualization of the degree distribution of nodes and their links, where the degree is represented by the size of the nodes (white circles).
	\begin{figure}[H]
		\centering
		\includegraphics[height=5.5cm,width=6cm]{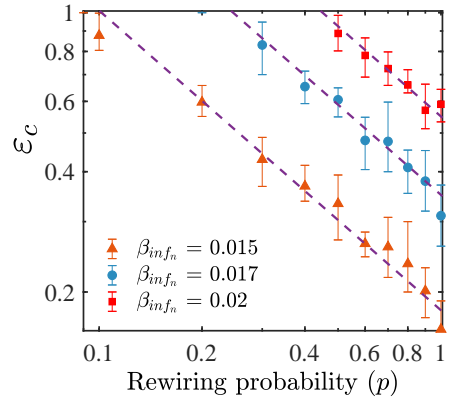}
		\caption{
			Critical  migration rate ($\epsilon_c$) against rewiring probability ($p$) plot  in log-log scale. Three different plots represent three forces of infection $\beta_{inf_n}$ = 0.015 (red triangle), 0.017 (blue circle), 0.02 (red square).  For each value of $p$, estimates of $\epsilon_c$ are averaged over 50 simulations on generated network samples and the variations  (standard deviation) are marked by bars. For each set of parameters ($\epsilon_c$, $p$ and $\beta_{inf_n}$), there is one emergent network that evolves with rewiring and becomes infection-free. The span of each plot decreases with higher $\beta_{inf_n}$ indicating that for a larger value, there exists no emergent network with a possibility of a disease-free state. All other parameters including $\beta_{o_m}=0.002$ are kept fixed as given in Fig.~\ref{fig2}.} \label{pe}
	\end{figure}
	\par We investigate the progress and recovery of infection with rewiring probability and its relation to the critical migration rate when a lattice is globally infection-free. For this purpose, for a fixed $\beta_{inf_n}$ and a given rewiring probability $p$, the critical migration rate $\epsilon=\epsilon_c$ is estimated. The estimation of $\epsilon_c$ is repeated with varying rewiring probability $p$. The estimation of $\epsilon_c$ with varying $p$ is then repeated for two additional $\beta_{inf_n}$ values.
	A plot of critical migration rate and rewiring probability follows a power law $\epsilon_c(p)$=$\kappa p^{-\alpha}$, as shown in Fig.~\ref{pe}. Numerically simulated data points  (triangles, circles and squares) of $\epsilon_c$ are plotted in a log-log scale 
	for three different  $\beta_{inf_n}$ = (0.015, 0.017, 0.02) and fitted with dashed lines that show an exponent value of  $\alpha=0.75$ with a normalization coefficient $\kappa$ set between $0.18\lesssim\kappa\lesssim0.5$. The statistical variation in the estimation of the critical migration rate for 50 repeat simulations is shown in bars in each plot for random variations in infected sites for each set of parameters. The power law explains that at a given $\beta_{inf_n}$ and for a lower rewiring probability $p$, a larger critical migration $\epsilon_c$ is necessary for a complete disease-free state with the emergence of a new network; increasing  $p$ can reduce the critical migration rate $\epsilon_c$ to a lower value for realizing a disease-free network. And for each set of parameters ($\epsilon_c$, $p$) there is one emergent network free from the disease. In principle, we can obtain a number of infection-free emergent networks for each set of given parameters. The possibility of getting any infection-free network decreases with increasing $\beta_{inf_n}$ values. An infection-free state is never achieved as revealed by the three plots for $\beta_{inf_n}$ = (0.015, 0.017, 0.02) where the span of the plots decreases with $\beta_{inf_n}$, and for larger values it vanishes.  
	An interesting point to note that the critical migration rates ($\epsilon_c\eqsim 0.185$, $0.33$, $0.63$) obtained by the power law at $p=1$ for the three cases are almost close (cf.  Fig.~\ref{all}(c)) to the critical rates of migration ($\epsilon_c$ = 0.2, 0.35, 0.65) obtained from the {\it RDIP} initial distribution for the same values of $\beta_{inf_n}$ = (0.015, 0.017, 0.02), which corresponds to global extinction of infection. It supports a general perception on the constructive role of random processes in many natural phenomena \cite{benzi1982stochastic,hanggi2002stochastic, van1994noise}.

	\section{Conclusion}
	Disease progression and recovery in a network is a complex process. We target a particular situation of disease spreading with a variation in the  initial pattern of infected sites in a network, and investigate how a particular pattern of initially infected sites affects the disease spreading with the dispersal of both infected and susceptible population, 
		and when a self-organized recovery process starts at all. We give due attention to the impact of the number of infected patches, the migration rate and the force of infection. The lattice is most vulnerable to infection if it is initiated from centrally located sites (${\it ICP}$ initial arrangement) and spreads via migration of both the infected and the susceptible population between the neighboring nodes. Global prevalence of the disease is most likely to occur if the initial number of infected patches and the migration rate are both high; no self-organized healing process starts except for a low force of infection. Disease management in such a case may be harder and controlling strategies should contain the restricted movement of the hosts and various disinfection methods. Comparably, spreading from initial infection in peripheral sites ({\it IPP case}), although increases with migration, shows signs of recovery at a higher rate of migration. On the contrary, an initial randomly distributed infected sites ${\it (RDIP)}$ in the lattice shows a higher resilience compared to other two initial arrangements ${\it ICP}$ and ${\it IPP}$. 
	In particular, the ${\it RDIP}$ initial distribution in the lattice plays a significant role in the recovery process besides its  dependence on the initial number of infected patches $(\%)$, rate of migration  and force of infection. 
	We estimated the largest initial fraction of infected patches $(\%)$ for the {\it RDIP} distribution that can achieve an infection-free state when the recovery process shows to follow an exponential law with relative infectivity. Seemingly, the initial ${\it ICP}$ arrangement is the worst performer both in the spreading and recovery process. In this particular case, a random rewiring of links or addition of long-range links as new migratory paths improves the disease state, and the global prevalence thereby declines and this condition progressively improves with the rewiring probability. Finally, at a critical migration rate and a rewiring probability, an infection-free network emerges. Interestingly, the critical migration rate and the rewiring probability that finally governs the recovery process  follows a power law. 
	In principle, for each set of ($\beta_{inf_n}$, $p$ and $\epsilon_c$) values, there is an emergent infection-free network, which, of course, does not exist only for larger $\beta_{inf_n}$. Although this study has not targeted a specific disease in the lattice, it explains the underlying mechanisms of disease spreading and the recovery process of many different types of diseases,  particularly, which are self-declining. While preparing our report, the world endemic arrives due to SARS-CoV-2, locking down the world over with the daily reports of new spread and death, which is still not under control and poorly understood. We do not rush to make any statement on this spreading and recovery of the disease at this time.
	
	\section*{Acknowledgment}
	This research is supported by the SERB (India), Grant No. EMR/2016/001078. C.H. is supported by DST-INSPIRE Faculty grant (code: IFA17-PH193).
	
	\section*{Appendix}
	
	In this Appendix, we elaborate that an arbitrary choice of $y_h$ threshold does not affect the qualitative nature of our main results. And then we show the temporal dynamics of all the infected and infection-free (susceptible) hosts in the whole population of all the sites in the lattice. They reveal their states for a long time run with increasing migration. 
	They present a real-time impression of the actual state of the whole population in all the sites how they evolve in time during a process of recovery or failure from recovery with increasing migration. The stability of the final stages of both the host populations (susceptible and infected) is analyzed using eigenvalue analysis of the lattice whose dynamics is defined by a set of coupled differential equations. A  semi-analytical technique has been adopted here for the coupled dynamical equations to derive the stability condition of the asymptotic states of the lattice. We form the {\it Jacobian} matrix of the coupled dynamical equations of the lattice by linearizing at the infection-free equilibrium state. And then numerically determine all the eigenvalues of the matrix. When the largest eigenvalue becomes negative or crosses the zero line (blue line) as shown in Figs.~\ref{all}(d)-\ref{all}(f), in the main text, with increasing migration $\epsilon$, we obtain a stable infection-free state. 
	
	\subsection{\bf Effect of the threshold  value $y_h$}
\begin{figure*}[!h]
		\centering
		\includegraphic[height=4cm,width=4.4cm]{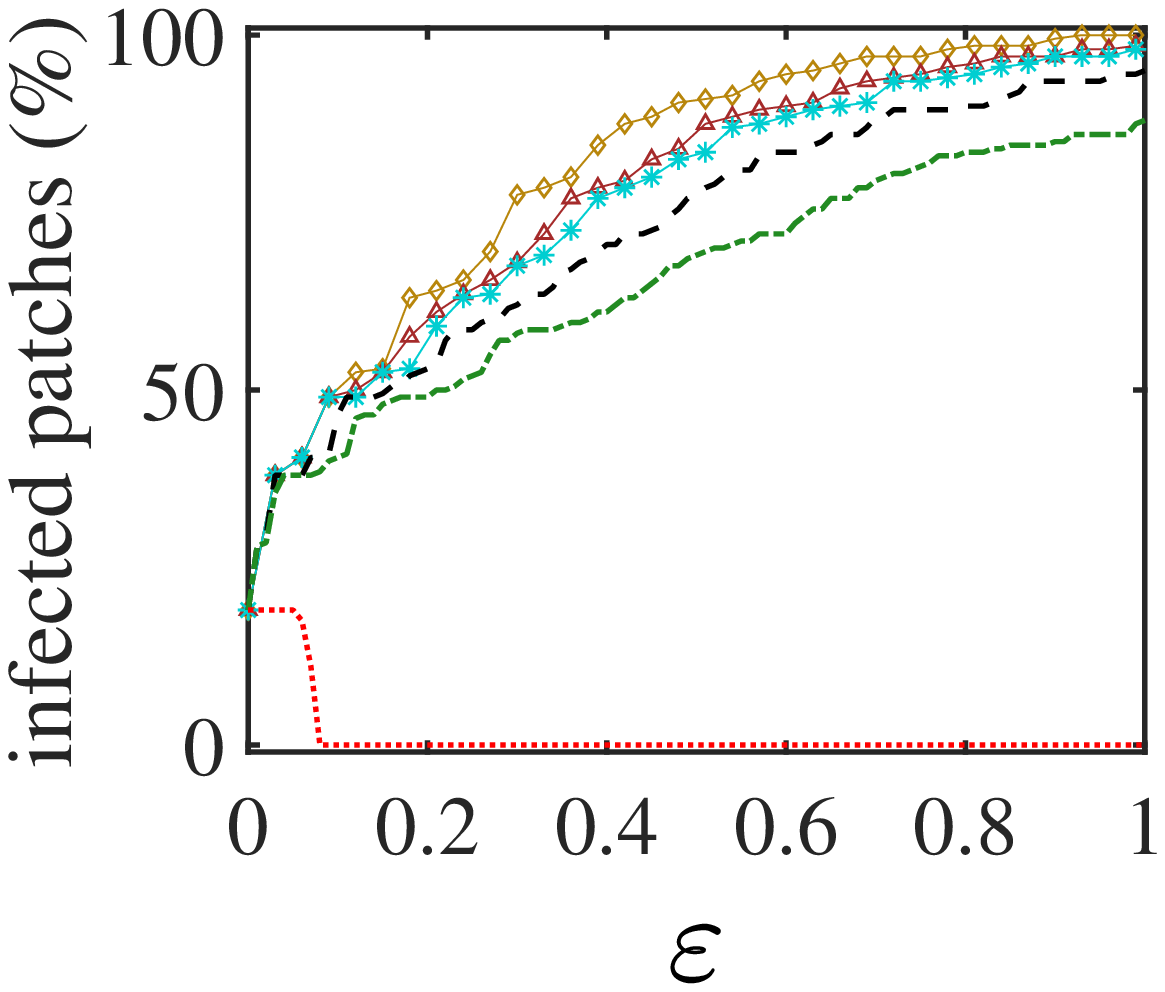}{(a) \quad~~  {\it ICP}}{35pt}{-0.0}
		\includegraphic[height=4cm,width=4.2cm]{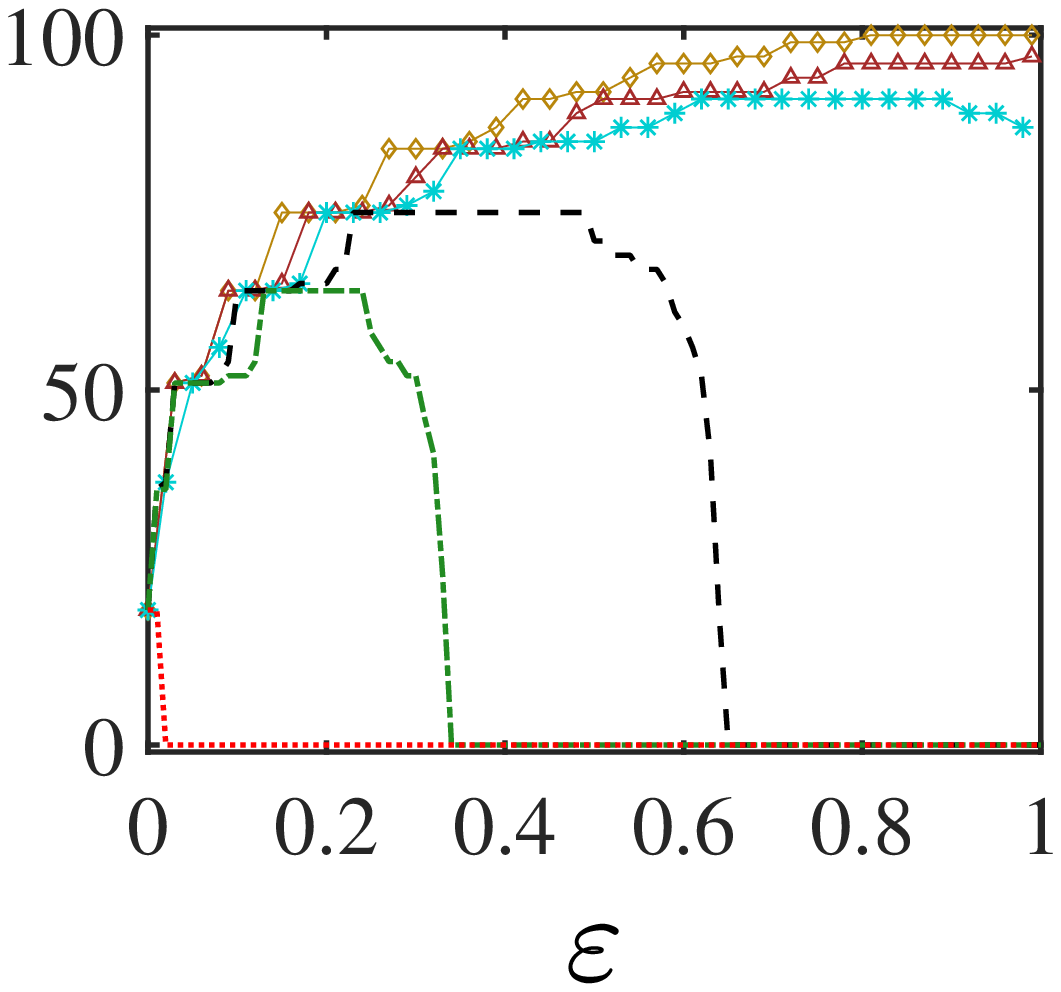}{(b)\quad \quad {\it IPP}}{25pt}{-0.0}
		\includegraphic[height=4cm,width=4.2cm]{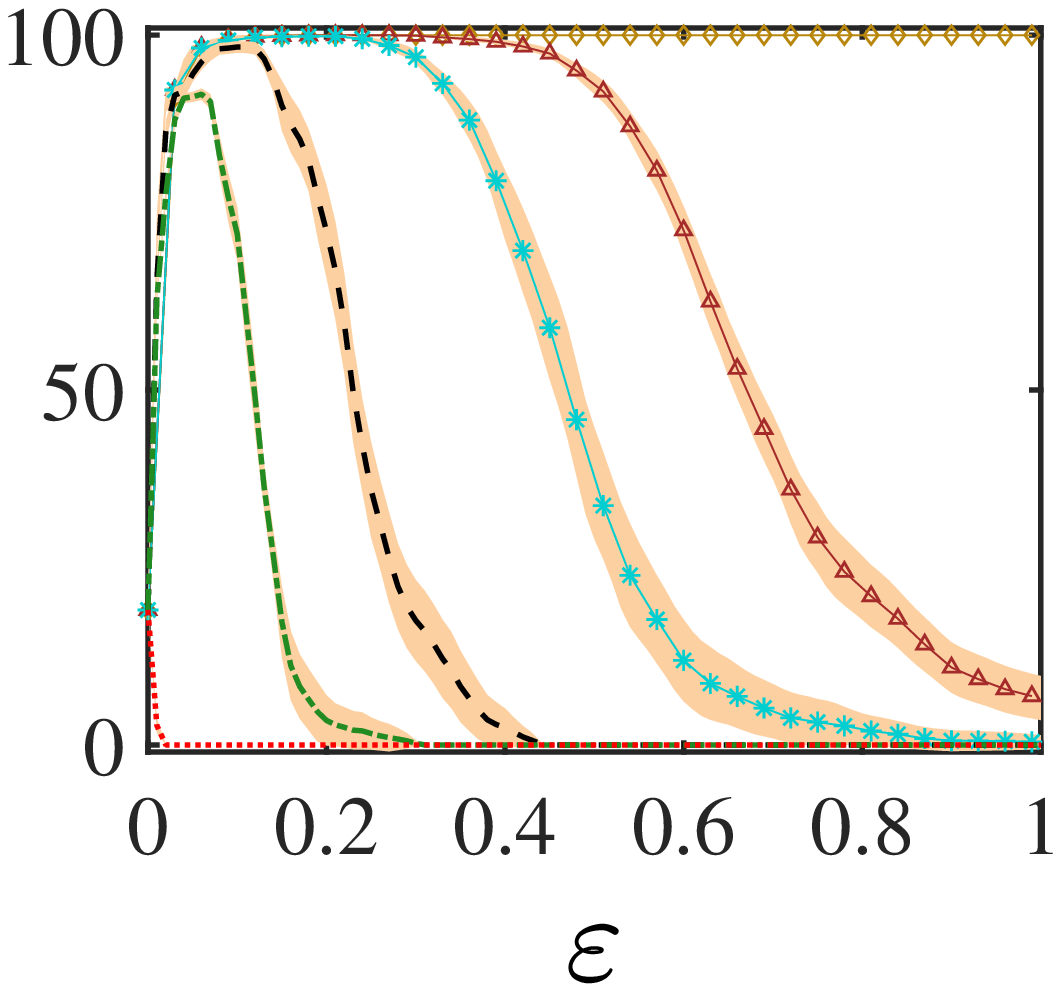}{(c)\quad \quad {\it RDIP}}{25pt}{-0.0} \\
		\includegraphics[height=0.25in,width=5in]{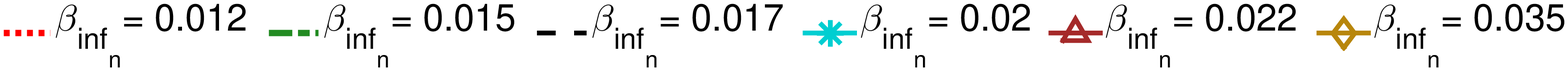}
		\caption{ Infection spreading and recovery process in the lattice with migration $\epsilon$. To define an infected patch, the threshold is chosen as $y_i>y_h=0.001$. Each of (a) {\it ICP}, (b) {\it IPP} and (c) {\it RDIP} case  starts with $19\%$ infected patches for five different force of infections $\beta_{inf_n}$=($0.012, 0.015, 0.017, 0.02, 0.022, 0.035$). Parameters are as in Fig.~\ref{all}. It shows that infection spreading and recovery is not affected by the considered threshold value, $y_h=0.01$. } \label{all2} 
	\end{figure*}

	In the main text, we chose the threshold value $y_h=0.01$ arbitrarily to define an infected patch and plotted the Figs.~\ref{all}(a)-\ref{all}(c) to show how infected patches recover from the disease with increasing migration for three considered initial arrangements of infected sites. We reproduced the results in Figs.~\ref{all2}(a)-\ref{all2}(c) here with a lower threshold $y_h=0.001$. It confirms that the qualitative nature of the results with $y_h=0.001$ remains the same with those of $y_h=0.01$. 
	
	\subsection{\bf Time evolutions of susceptible and infected population}
\begin{figure*}[!h]
		\centering
		\includegraphic[height=2.5cm,width=4cm]{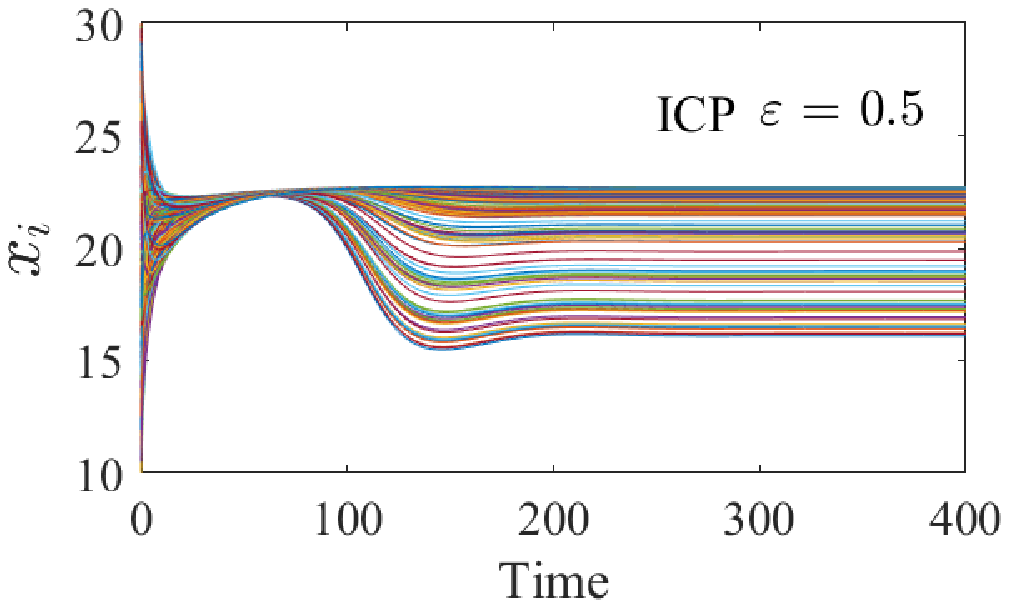}{(a)}{20pt}{1}
		\includegraphic[height=2.5cm,width=4cm]{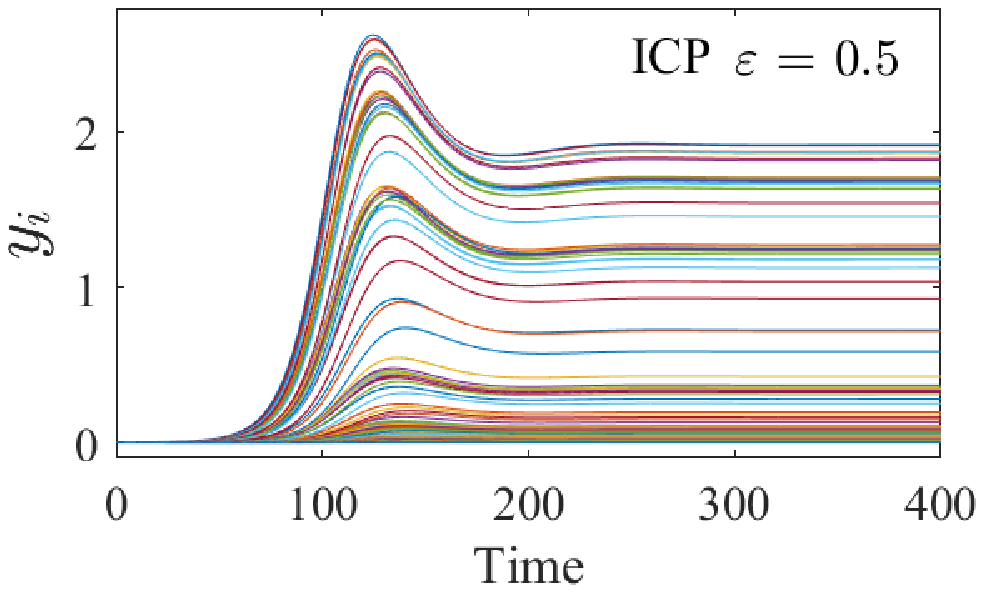}{(b)}{20pt}{1}
		\includegraphic[height=2.5cm,width=4cm]{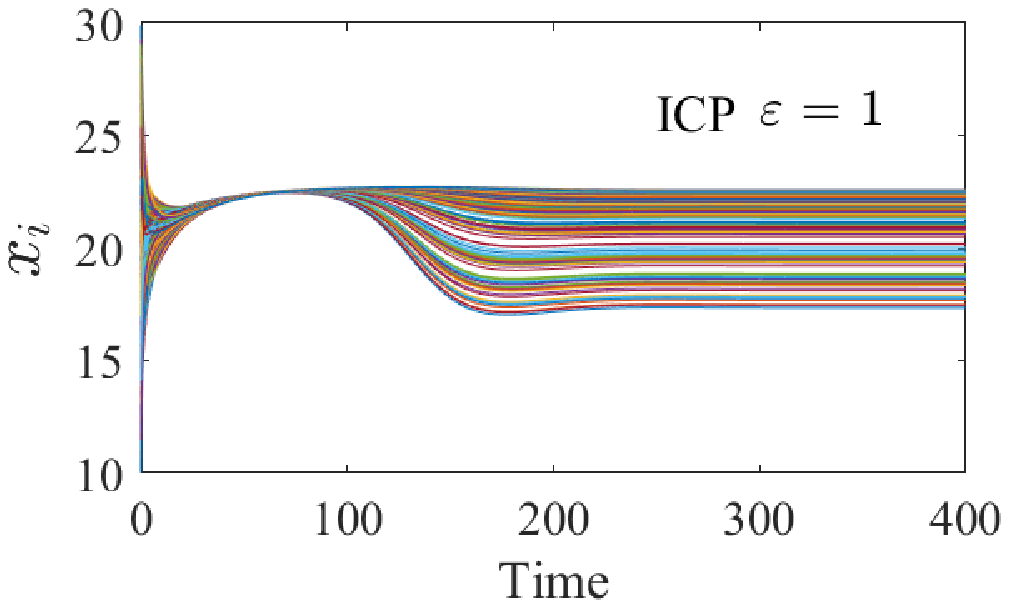}{(c)}{20pt}{1}
		\includegraphic[height=2.5cm,width=4cm]{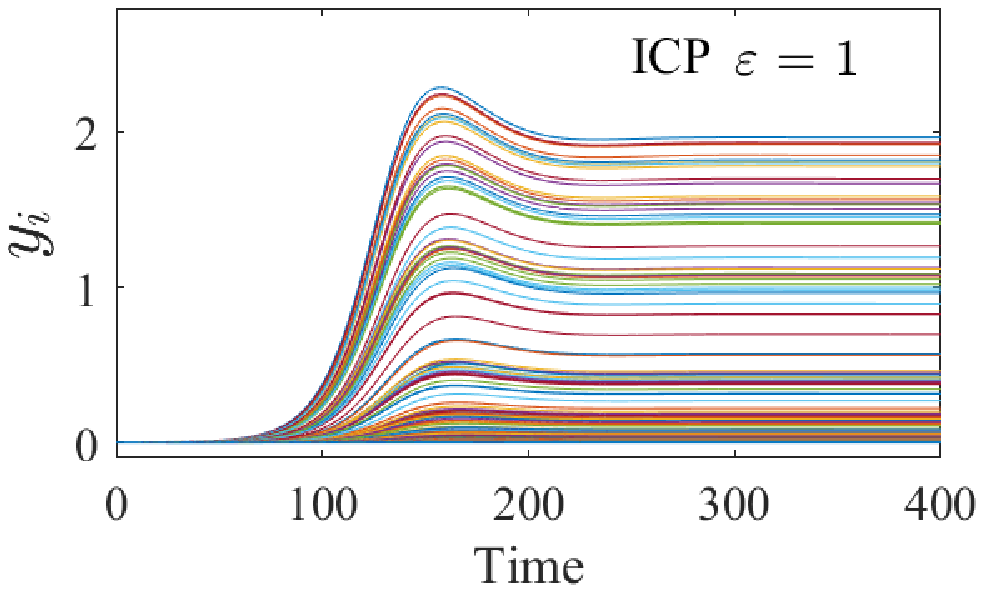}{(d)}{20pt}{1}\\
		\includegraphic[height=2.5cm,width=4cm]{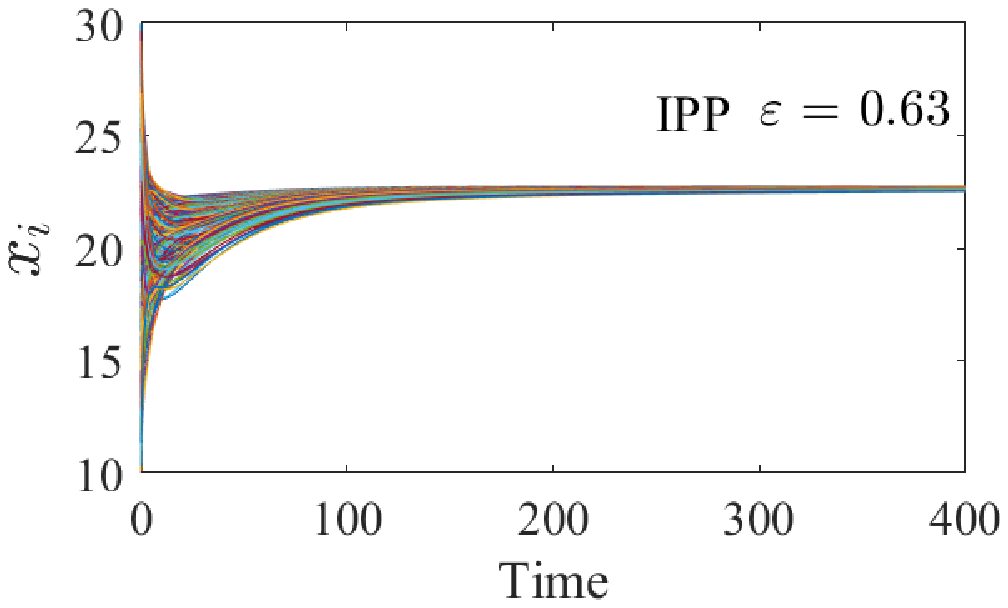}{(e)}{20pt}{1}
		\includegraphic[height=2.5cm,width=4cm]{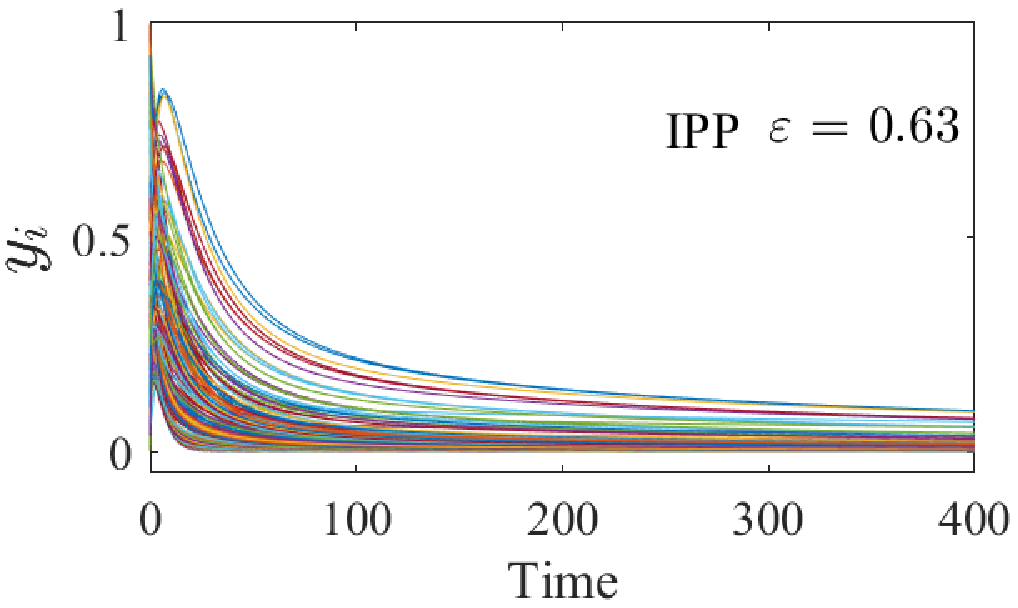}{(f)}{20pt}{1}
		\includegraphic[height=2.5cm,width=4cm]{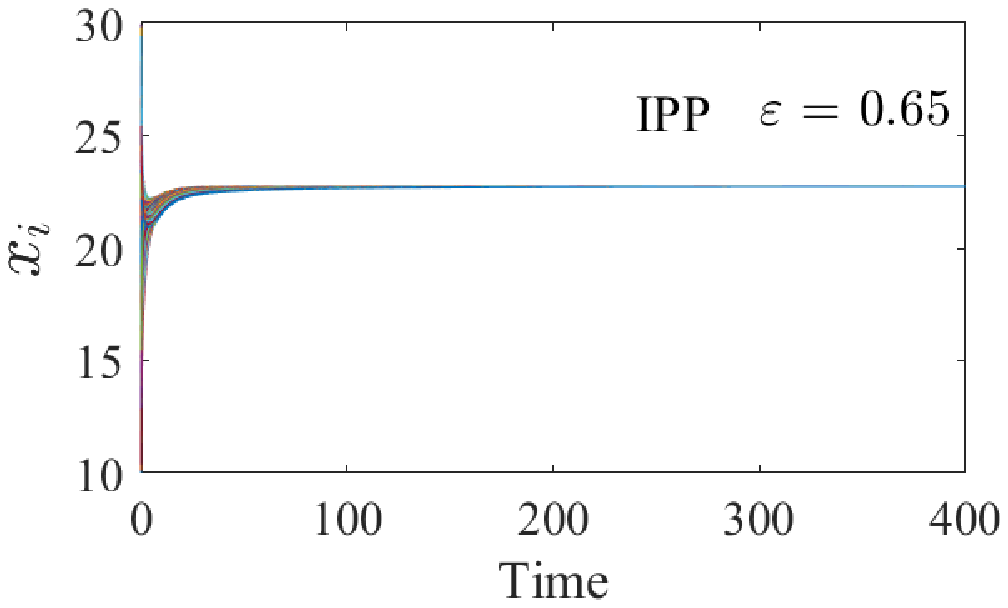}{(g)}{20pt}{1}
		\includegraphic[height=2.5cm,width=4cm]{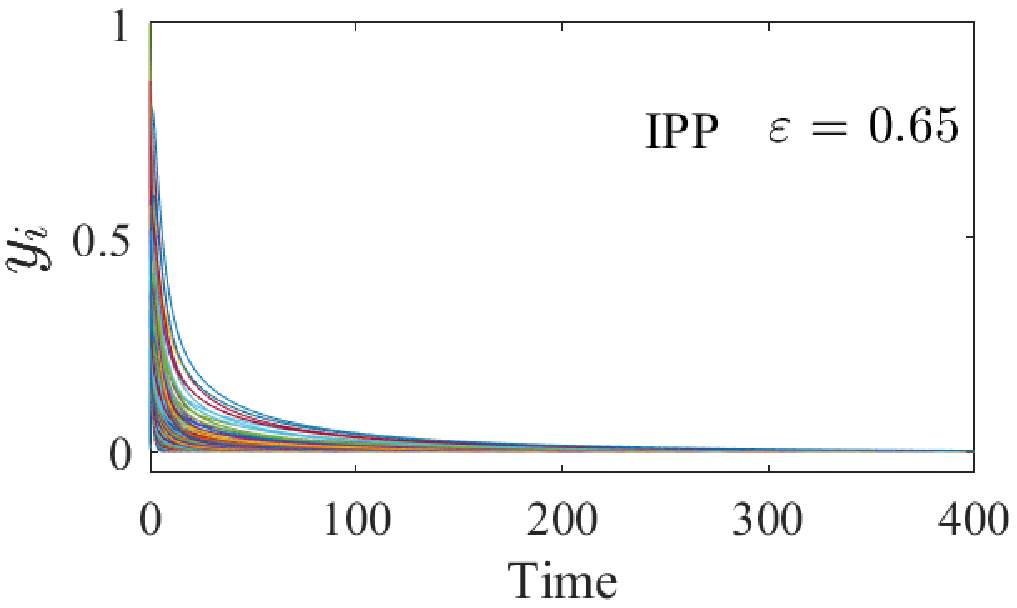}{(h)}{20pt}{1}\\
		\includegraphic[height=2.5cm,width=4cm]{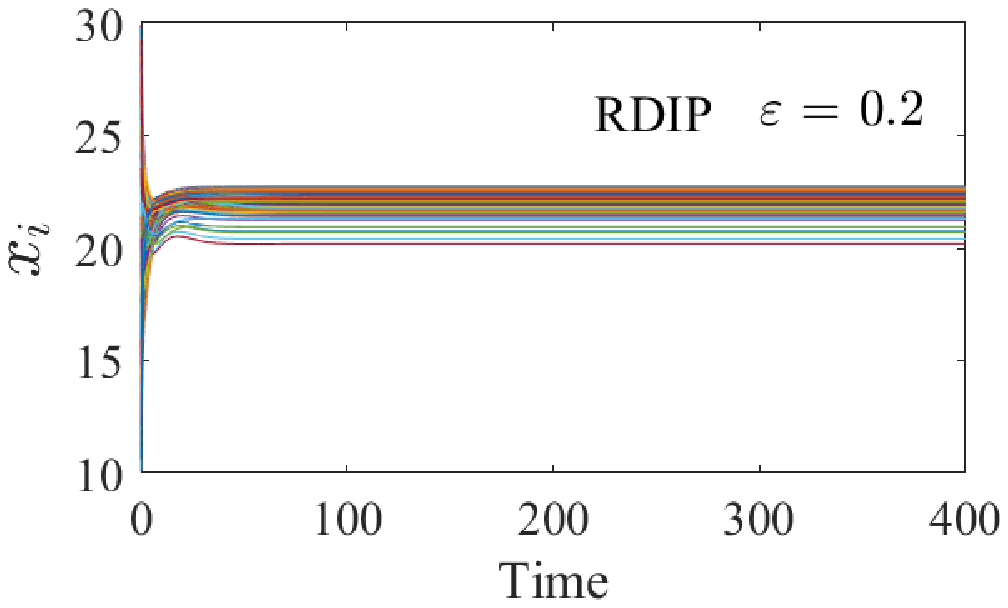}{(i)}{20pt}{1}
		\includegraphic[height=2.5cm,width=4cm]{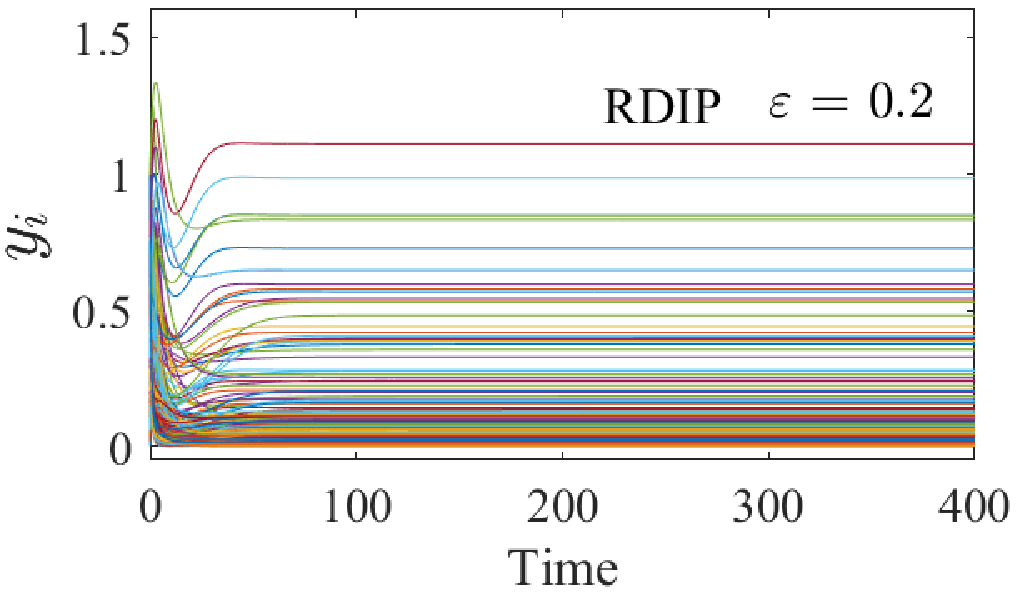}{(j)}{20pt}{1}
		\includegraphic[height=2.5cm,width=4cm]{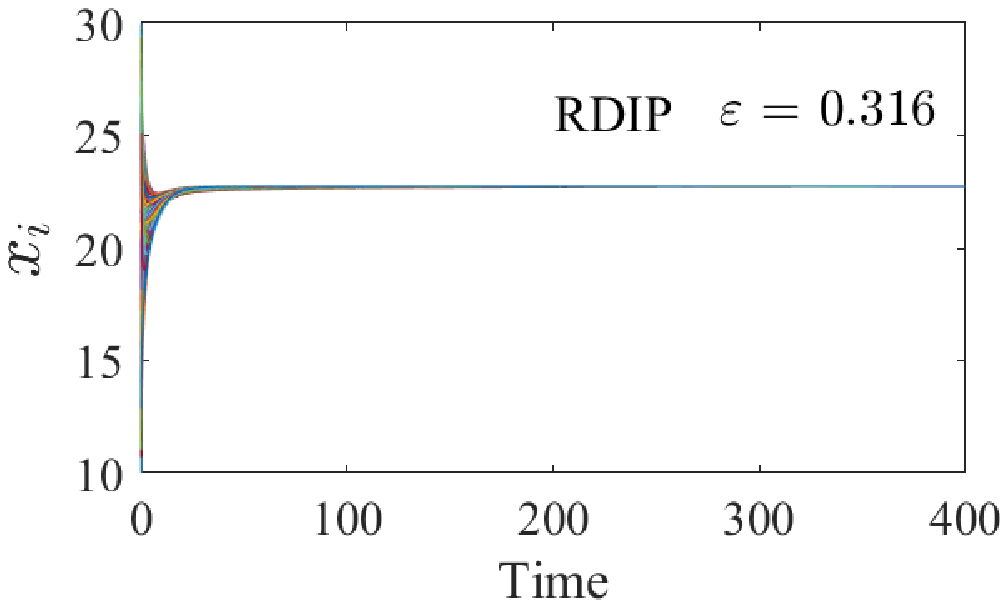}{(k)}{20pt}{1}
		\includegraphic[height=2.5cm,width=4cm]{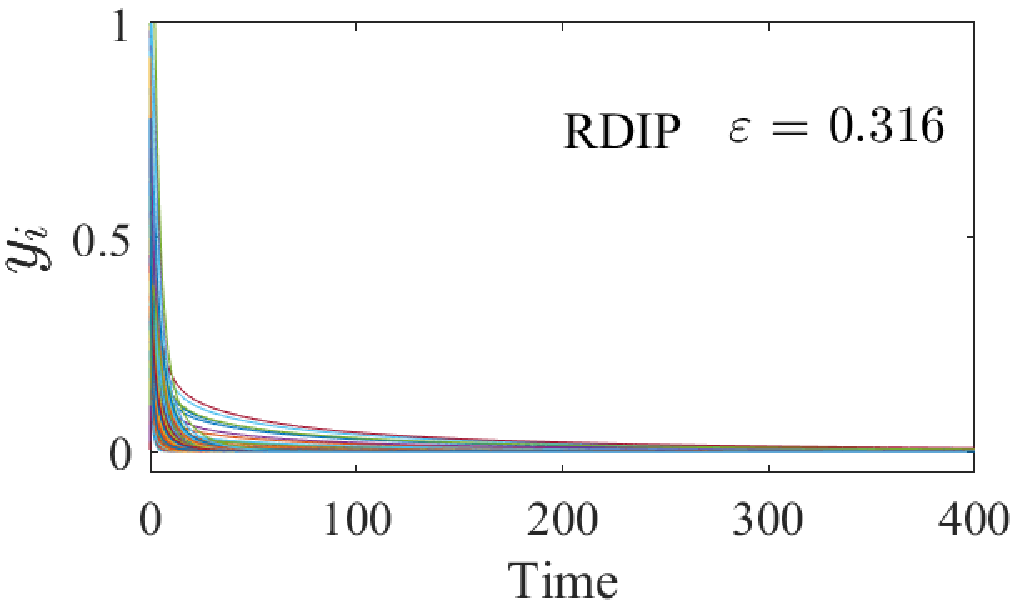}{(l)}{20pt}{1}
		\caption{Temporal evolution of susceptible ($x_i$) and infected ($y_i$) population. First row: Population densities of the lattice with {\it ICP} arrangement for $\epsilon =0.5$ (a-b) and $\epsilon =1$ (c-d) corresponding to Figs.~\ref{snap_grid}(A4) and \ref{snap_grid}(A5), respectively. These figures show that infection persists in the lattice for both the cases. Second row: Population densities of the lattice with {\it IPP} arrangement for $\epsilon =0.63$ (e-f) and $\epsilon =0.65$ (g-h) corresponding to Figs.~\ref{snap_grid}(B4) and \ref{snap_grid}(B5), respectively. These figures show that infection persists in the lattice for $\epsilon =0.63$ but dies out for $\epsilon =0.65$. Third row: Population densities of the lattice with {\it RDIP} arrangement for $\epsilon =0.2$ (i-j) and $\epsilon =0.316$ (k-l) corresponding to Figs.~\ref{snap_grid}(C4) and \ref{snap_grid}(C5), respectively. These figures show that infection persists in the lattice for $\epsilon =0.2$ but dies out for $\epsilon =0.316$. Parameters are as in Fig.~\ref{snap_grid} }. \label{time_ABC45}
	\end{figure*}

	Here we show the temporal evolutions of the susceptible ($x_i$) and infected hosts ($y_i$) in all the sites in Fig.~\ref{time_ABC45} with increasing migration that correspond to Fig.~\ref{snap_grid}. Such a description of time evolution picturizes the real time evolution of each host in all sites. For the {\it ICP} case, time evolutions of $x_i$ and $y_i$ are shown in Figs.~\ref{time_ABC45}(a) and \ref{time_ABC45}(b) for $\epsilon=0.5$ (corresponds to panel A4 in Fig.~\ref{snap_grid}), which confirms  that the susceptible population takes different stable values ($x_i \neq 0$) in the long time run while the infected population also assumes different values ranging from $y_i=0$ to $y_i\neq 0$. This particular state is defined as the IHSS in the literature \cite{hens2013oscillation, koseska2013transition, hens2014diverse, nandan2014transition}. Some more patches are infected when the migration rate becomes higher as shown in their time evolutions in Figs.~\ref{time_ABC45}(c) and \ref{time_ABC45}(d) for $\epsilon=1.0$ (cf. panel A5 in Fig.~\ref{snap_grid}). 
	The IHSS state explains the coexistence of susceptible and infected population in long time, which never converge to a unique infection-free stable state ($x_i\neq 0$, $y_i=0$). In the case of {\it IPP}, IHSS has also emergeed as shown for a migration rate $\epsilon=0.63$  in Figs.~\ref{time_ABC45}(e) and \ref{time_ABC45}(f) (cf. panel B4 in Fig.~\ref{snap_grid}). However, the steady state values are slowly converging, and finally reaches the HSS state for $\epsilon=0.65$; time evolutions of $x_i$ and $ y_i$ in Figs.~\ref{time_ABC45}(g) and \ref{time_ABC45}(h) (cf. panel B5 in Fig.~\ref{snap_grid}) show that both the hosts converge to the unique infection-free equilibrium state ($x_i\neq 0, y_i=0$). Such a unique equilibrium state as reached by all the hosts in the lattice is defined as the homogeneous steady state (HSS) in the literature \cite{hens2013oscillation, koseska2013transition, hens2014diverse, nandan2014transition}. This HSS basically represents a stable equilibrium state of an uncoupled node [$E_s=({x_s,0})$], where the entire lattice is infection-free, as confirmed by our eigenvalue analysis. For the {\it RDIP} case, the time evolution of the hosts $x_i$ and $ y_i$ passes through an IHSS state for an intermediate value of $\epsilon=0.2$ as shown in Figs.~\ref{time_ABC45}(i) and \ref{time_ABC45}(j) (cf. panel C4 in Fig.~\ref{snap_grid}) and eventually converge to the unique HSS in Figs.~\ref{time_ABC45}(k) and \ref{time_ABC45}(l) (cf. panel C5 in Fig.~\ref{snap_grid}) for $\epsilon=0.316$.  
	
	\subsection{\bf Stability of the Infection-free equilibrium of the lattice}
	
	In the infection-fee state, the lattice reaches an HSS state, when the susceptible population arrives at the stable equilibrium $E_s(x_s, 0)$ where infected population no more exists $y_i=0$. A semi-analytical strategy is performed to confirm the local stability of the HSS state first by linearizing the coupled equations  \eqref{eq2} representing the lattice, to derive the {\it Jacobian} matrix at the infection-free equilibrium $E_s=(x_s, 0)$, then numerically calculating all the eigenvalues of the matrix.
	The {\it Jacobian} matrix of the coupled system of the square lattice of $20\times20$ patches  \eqref{eq2} at the infection-free equilibrium $E_s=(x_s, 0)$ then reads
	\begin{widetext}
		{ \scriptsize
			\begin{align}
			{\bf J_{(x_s,0)}=}\begin{bmatrix}
			\frac{x_sb_1(2c_1-1)}{c_1(1-c_1)}-d_1-\epsilon,& \frac{x_sb_1(2c_1-1)}{c_1(1-c_1)}+\gamma-x_s\beta_1,& \frac{\epsilon}{2},& 0,& 0,& 0, \cdots \; 0,& 0,& 0,& 0 \\
			0,& x_s\beta_1-d_1-\gamma-\alpha-\epsilon,& 0,& \frac{\epsilon}{2},& 0,& 0, \cdots \; 0,& 0,& 0,& 0 \\
			\frac{\epsilon}{3},& 0,& \frac{x_sb_1(2c_1-1)}{c_1(1-c_1)}-d_1-\epsilon,& \frac{x_sb_1(2c_1-1)}{c_1(1-c_1)}+\gamma-x_s\beta_{2},& \frac{\epsilon}{3},& 0, \cdots \; 0,& 0,& 0,& 0 \\
			0,& \frac{\epsilon}{3},& 0,& x_s\beta_2-d_1-\gamma-\alpha-\epsilon,& 0,&  \frac{\epsilon}{3}, \cdots \; 0,& 0,& 0,& 0 \\
			\vdots&&&&\ddots&&&&\vdots\\
			\vdots&&&&&\ddots&&&\vdots\\ 
			0,& 0,& 0,& 0,& 0,& 0, \cdots \; \frac{\epsilon}{2},& 0,& \frac{x_sb_1(2c_1-1)}{c_1(1-c_1)}-d_1-\epsilon,& \frac{x_sb_1(2c_1-1)}{c_1(1-c_1)}+\gamma-x_s\beta_{400} \\
			0,& 0,& 0,& 0,& 0,& 0, \cdots \; 0,& \frac{\epsilon}{2},& 0,& x_s\beta_{400}-d_1-\gamma-\alpha-\epsilon
			\end{bmatrix}. \label{jac2}
			\end{align} }	
	\end{widetext}
	
	\noindent {We numerically calculate the eigenvalues of ${\bf J_{(x_s,0)}}$ with varying $\epsilon$. Let, the maximum of the real eigenvalue spectrum of the {\it Jacobian} \eqref{jac2} at $E_s=(x_s,0)$ as a function of $\epsilon$ is defined as 
		\begin{align}\label{eq5}
		\lambda_{max}(\epsilon)=Re\left| eig[J_{(x_s,0)}]\right|_{max}.
		\end{align}
		The critical value, $\epsilon_c$, is numerically obtained when $\lambda_{max}$ becomes negative from its positive value. Thus, 
		\begin{align}
		\lambda_{max}(\epsilon)<0, \quad \text{for} \quad \epsilon>\epsilon_c. \label{cond}
		\end{align}
		Under the condition \eqref{cond}, infection-free equilibrium state is locally stable, implying that the entire lattice  recovers completely from the infected state. The largest eigenvalues for HSS are plotted in Figs.~\ref{all}(g)-\ref{all}(i), in the main text. Thereby, semi-analytically estimated the critical values $\epsilon=\epsilon_c$ for different $\beta_{inf_n}$ in the cases of {\it ICP}, {\it IPP}, {\it RDIP} arrangements and are verified with their numerical counterparts shown in Figs.~\ref{all}(a)-\ref{all}(c). 
		The critical migration rates are estimated for all the three cases: (I) {\it ICP} initial arrangement, $\epsilon_c=0.07$ for $\beta_{inf_n}=0.012$ (only one case when the lattice becomes infection-free), (II) {\it IPP} arrangement, $\epsilon_c$=0.02, 0.34 and 0.65 for $\beta_{inf_n}$=0.012, 0.015 and 0.017, respectively. 
		(III) {\it RDIP} arrangement, $\epsilon_c$=0.01, 0.2, 0.35 and 0.66 for $\beta_{inf_n}$=0.012, 0.015, 0.017 and 0.02, respectively.

\bibliography{ref-final}
\end{document}